\documentclass[manuscript]{acmart}

\AtBeginDocument{%
  }

\usepackage{braket}
\usepackage{physics}
\usepackage{titlesec}
\usepackage{listings}
\usepackage[noend]{algpseudocode}
\usepackage{algorithm}
\usepackage{hyperref}
\usepackage{url}
\usepackage{xcolor}
\usepackage{color}
\usepackage{xspace}
\usepackage{multicol}
\usepackage{tabularx}
\usepackage{graphicx}
\usepackage{lscape}
\usepackage{tikz}
\usetikzlibrary{quantikz2}
\usepackage{multirow}

\definecolor{backcolor}{rgb}{0.95,0.95,0.92}
\definecolor{codegray}{rgb}{0.5,0.5,0.5}
\definecolor{codegreen}{rgb}{0,0.6,0}
\definecolor{codeblue}{rgb}{0.015,0.015,0.66}
\definecolor{codepurple}{rgb}{0.58,0,0.82}
\definecolor{codered}{rgb}{1.0, 0.0, 0}

\tikzset{
    operator/.append style={fill=blue!20,  color=blue!60},
}

\newcommand{\vcwdouble}[3]{
    \arrow[from=#1,to=#2,arrows, Rightarrow, double distance=.2mm, line width=.2mm, ->, shorten >=1pt, >=latex, "#3" {anchor=south west, yshift=1pt,xshift=2pt},  at end ] 
}
\begin{document}

\title{Testing and Debugging Quantum Programs: The Road to 2030}

\author{Neilson C. L. Ramalho}
\email{neilson@usp.br}
\orcid{0009-0007-8543-6426}
\affiliation{%
  \institution{School of Arts, Sciences, and Humanities -- University of São Paulo}
  \city{São Paulo}
  \state{São Paulo}
  \country{Brazil}
  \postcode{03828-000}
}
\author{Higor A. de Souza}
\email{higor.souza@unesp.br}
\orcid{0000-0003-4233-5987}
\affiliation{%
  \institution{Department of Computing -- São Paulo State University}
  \city{Bauru}
  \state{São Paulo}
  \country{Brazil}
  \postcode{17033-360}
}

\author{Marcos Lordello Chaim}
\email{chaim@usp.br}
\orcid{0000-0001-7157-5141}
\affiliation{%
  \institution{School of Arts, Sciences, and Humanities -- University of São Paulo}
  \city{São Paulo}
  \state{São Paulo}
  \country{Brazil}
  \postcode{03828-000}
}
\renewcommand{\shortauthors}{Ramalho et al.}

\begin{abstract}
Quantum computing has existed in the theoretical realm for several decades. Recently, quantum computing has re-emerged as a promising technology to solve problems that a classical computer could take hundreds of years to solve. However, there are challenges and opportunities for academics and practitioners regarding software engineering practices for testing and debugging quantum programs. This paper presents a roadmap for addressing these challenges, pointing out the existing gaps in the literature and suggesting research directions. We discuss the limitations caused by noise, the no-cloning theorem, the lack of a standard architecture for quantum computers, among others. Regarding testing, we highlight gaps and opportunities related to transpilation, mutation analysis, input states with hybrid interfaces, program analysis, and coverage. For debugging, we present the current strategies, including classical techniques applied to quantum programs, quantum-specific assertions, and quantum-related bug patterns. We introduce a conceptual model to illustrate concepts regarding the testing and debugging of quantum programs and the relationship between them. Those concepts are used to identify and discuss research challenges to cope with quantum programs through 2030, focusing on the interfaces between classical and quantum computing and on creating testing and debugging techniques that take advantage of the unique quantum computing characteristics.
\end{abstract}



\maketitle

\section{Introduction}

Quantum computing has leaped forward in recent years, gaining attention from both academia and industry. The main reason for developing software and hardware solutions based on the quantum realm is the need to speed up the processing of complex problems.

A quantum computer is a device that takes advantage of the specific properties described by quantum mechanics to perform computation \cite{Hidary2019Quantum}. While quantum computers have been a theoretical concept for years, several companies are now engaged in developing hardware, programming frameworks, and quantum programming languages such as Q\# from Microsoft\footnote{\href{https://learn.microsoft.com/en-us/azure/quantum/user-guide/}{https://learn.microsoft.com/en-us/azure/quantum/user-guide/}}, Cirq from Google\footnote{\href{https://quantumai.google/cirq}{https://quantumai.google/cirq}}, and Qiskit from IBM\footnote{\href{https://qiskit.org/}{https://qiskit.org/}}. As hardware development progresses, quantum computing applications become increasingly important and promising, with potential in fields such as molecular simulations, cybersecurity, finance, and logistics. Quantum computing has also been used to accelerate the execution of classical machine learning and to create new quantum machine learning algorithms \cite{combarro2022}. 

Software Engineering practices and techniques must support the development of quantum applications to achieve productivity, quality, and business-oriented solutions. On one hand, software engineers must know the fundamental basis of quantum computing to understand its specificities. On the other hand, specialists in developing quantum applications should comprehend the importance of software engineering practices, methods, and techniques to deliver bug-free applications that can be maintained during their lifecycles. 

With the spread of frameworks and programming languages, it is important to ensure that quantum computing applications will work as expected, both as standalone components and as sub-modules of larger hybrid applications with classical computing components. Quantum computing has certain characteristics that pose new challenges for testing and debugging for researchers and practitioners \cite{stancil2022principles}. For instance, in the current Noisy Intermediate-Scale Quantum (NISQ), quantum computers are still constrained by noise, limiting their reliability for large-scale computations \cite{preskill2018nisq}.

To keep up with advances made by practitioners, researchers are proposing strategies to test and debug quantum applications, adapting existing techniques or creating new approaches based on quantum computing concepts.

This paper extends the work presented at the ``2030 Software Engineering'' workshop, held during the ACM International Conference on the Foundations of Software Engineering (FSE 2024). It presents a roadmap with insights concerning the future of testing and debugging of quantum programs. Compared to the previous paper, this version includes the following enhancements.
\begin{enumerate} 
    \item An extended overview of Quantum Machine Learning (QML) encoding techniques, given their importance as the interfaces between classical and quantum data. 
    \item A detailed discussion on quantum assertions. 
    \item Inclusion of recent literature in areas such as flaky tests, transpilation, code smells, bug patterns, and program analysis. 
    \item An updated version of the conceptual model, presenting new concepts and relevant relations to support the understanding of the field.
    \item Timelines describing the evolution of testing and debugging techniques for quantum programs.
    \item An expanded discussion on research directions, with the identification of 12 challenges followed by possible research opportunities, and new areas of exploration. 
\end{enumerate}

We will first provide a brief theoretical background about quantum computing in Section~\ref{sec:quantum-computing}. Concepts such as quantum bits (qubits), superposition, and entanglement are explained and illustrated with an example. By doing so, we will be able to discuss the impact of those quantum characteristics for testing and debugging. 
In Section~\ref{sec:current-approaches}, we present an overview of the state-of-the-art and state-of-the-practice in quantum computing, such as classical testing and debugging techniques applied for quantum programs, quantum assertions, bug patterns, and bug hierarchies in quantum programs.
In Section~\ref{sec:conceptual-model}, we propose a conceptual model that represents the topics related to testing and debugging quantum applications. This model will be used to discuss concerns and challenges that may be addressed by researchers for the future on the road to 2030 in Section~\ref{sec:challenges}. Section~\ref{sec:challenges} also summarizes the key research opportunities we envision for the coming years. Section~\ref{sec:related-work} presents the related work and Section~\ref{sec:conclusions} contains our remarks and conclusions.

\section{Quantum computing concepts}
\label{sec:quantum-computing}
To understand the differences between a Classical Program (CP) and a Quantum Program (QP) in terms of testing and debugging, it is useful to explore some of the key characteristics of QC \cite{stancil2022principles}:

\begin{itemize}
	\item \textit{Quantum parallelism}: Superposition is the quantum principle that allows calculations to consider multiple quantum states simultaneously, thus leveraging parallelism. 
 
	\item \textit{Statistical results in most cases}: unlike classical programs, most quantum computing applications are governed by the inherent uncertainty of superposition. 
 
	\item \textit{Exponential scaling}:  as qubits can assume two values ($\ket{0}$ and $\ket{1}$), the input space of quantum programs scales as $2\textsuperscript{N}$, where N is the number of qubits. With 50 qubits, for example, the number of possibilities increases to $2\textsuperscript{50}$, which can not be simulated even by current supercomputers. 
	
	\item \textit{Quantum interference}:  interference occurs when two quantum states are combined so that their amplitudes either amplify (constructive interference for the right answer) or cancel (destructive interference for incorrect answers).
	
	\item \textit{Asking the right question}: This characteristic refers to the mapping of the problem statement such as it can be solved by a quantum computer.

\end{itemize}
Besides the concepts mentioned above, there are also characteristics such as entanglement and the no-cloning theorem that directly impact the testing and debugging of quantum programs. These concepts will be discussed in detail in the following paragraphs with the support of the running example of the Bell state circuit presented in Listing \ref{alg:algoritmo-running-example1}. This circuit uses two qubits to generate four maximally entangled states.

\begin{lstlisting}[language=Python, label={alg:algoritmo-running-example1},caption=Running example with the Bell state circuit]
from qiskit import QuantumCircuit
from qiskit_aer import Aer
from qiskit import transpile
from qiskit.visualization import plot_histogram
from matplotlib import pyplot as plt

# creating a circuit with 2 classical
# registers and 2 quantum registers
circuit = QuantumCircuit(2)

circuit.h(0)
circuit.cx(0,1)
circuit.measure_all()

# Transpile for simulator
simulator = Aer.get_backend('aer_simulator')
circuit = transpile(circuit,simulator)
# simulates the circuit
result = simulator.run(circuit,shots=10000).result()
counts = result.get_counts(circuit)
plot_histogram(counts)
plt.show()
\end{lstlisting}

The code is written in Python and is based on Qiskit, an open-source software development kit for working with quantum computers at the level of pulses, circuits, and application modules \cite{qiskit2024}. One of the central aspects of programming using Qiskit is the circuit, which consists of one or more quantum operators. Figure \ref{fig:running_example} presents the code for the quantum circuit described in Listing \ref{alg:algoritmo-running-example1}. Quantum circuits are composed of wires and quantum gates, responsible for carrying around and manipulating quantum information \cite{NielsenMichaelChuang}. The vertical dashed lines are called barriers and are used in this example to divide the circuit into pieces so it is easier to explain the parts individually. The circuit of Figure \ref{fig:running_example} is composed of two qubits ($q_0$ and $q_1$), a Hadamard, and a CNOT operator (gate), as well as the measurements for each qubit. Table \ref{tab:main_quantum_gates} contains a summary of some of the main quantum operators (gates).

\begin{figure}[ht]
\begin{center}
    \begin{quantikz}[wire types={q,q,c}]
        \lstick{q$_0$} & \gate{H} & \ctrl{0} \vqw{1} & \meter{} \vcwdouble{1-4}{3-4}{0}   &       & \\
        \lstick{q$_1$} &          & \targ{}         &       & \meter{} \vcwdouble{2-5}{3-5}{1} &  \\  
        \lstick{c}     & \qwbundle{2}      &             &       &       &  
    \end{quantikz}
\end{center}
  \caption{Running example circuit}
  \label{fig:running_example}
\end{figure}

Similarly to classical bits, qubits can assume two different measurable states: $\ket{0}$ and $\ket{1}$, which are, to a certain extent, equivalent to the classical binary states 0 and 1. However, since qubits are physically subatomic particles, they have certain quantum mechanical properties such as superposition of states and entanglement. Computations that would normally need to be performed serially on 0 and 1 separately on a classical computer could now be completed in a single operation using a qubit on a quantum computer, making computations faster \cite{Hughes2021Quantum}. 

Qubits can be in different states, represented in Quantum Mechanics as vectors, usually with the Dirac notation. In this notation, vectors are represented as bra-kets, in which a ket is the column vector and a bra its conjugate transpose. In this way, the vectors representing the states $\ket{0}$ and $\ket{1}$ are defined as:

$$
\ket{0} = 
\begin{pmatrix}
1 \\
0
\end{pmatrix}, \; \;
\ket{1} = 
\begin{pmatrix}
0 \\
1
\end{pmatrix}
$$

In Listing \ref{alg:algoritmo-running-example1}, line 9, the constructor \textit{QuantumCircuit} creates two qubits initialized with the base state $\ket{0}$ and two classical registers to store the results of the measurements operations in each qubit created. 
The circuit is composed of $q_0$ and $q_1$ and is, at this point, a composite system. A system with multiple qubits is called a composite quantum system \cite{laforest2015mathematics} and is the result of the tensor product of the separate, individual spaces. For instance, a composite system with two qubits consists of a single quantum system with four dimensions. In general terms, a composite of a quantum system of N qubits consists of a single quantum system with $ 2^N $ dimensions.

The composition is represented by the symbol $ \otimes $, which is the tensor product of the state vectors that represent each qubit individually. For two qubits, the computational basis is $\ket{00}$, $\ket{01}$, $\ket{10}$, and $\ket{11}$. The tensor product is defined in the Dirac notation as:
$$ 
\ket{x} \otimes \ket{y} \equiv \ket{x} \ket{y} \equiv \ket{xy}
$$
As stated by quantum mechanics principles, systems are set to a definite state only once they are measured. Before a measurement, the systems are in an indeterminate state. For instance, the superposition of $\ket{0}$ and $\ket{1}$ is a linear combination of these states:
\begin{equation}
  \ket{\psi} =  \frac{1}{\sqrt{2}}\ket{0} + \frac{1}{\sqrt{2}}\ket{1} 
	\label{eq:equacao1}
\end{equation}
The state presented in Equation \ref{eq:equacao1} is usually represented as $\ket{+}$, whereas the state $\ket{\psi} =  \frac{1}{\sqrt{2}}\ket{0} - \frac{1}{\sqrt{2}}\ket{1}$ is represented by $\ket{-}$.

When quantum states are in a superposition, the probability of a state resulting after the measurement is equal to the modulus squared of the amplitude of that state. This is known as the Born Rule and it was demonstrated by Max Born in 1926 \cite{Hidary2019Quantum}. In Equation \ref{eq:equacao1}, the probability of the state $\ket{0}$ or $\ket{1}$ being returned after measurement is equal to:
\begin{equation}
 P(\ket{0}) = P(\ket{1}) =  \abs{\frac{1}{\sqrt{2}}}^2 = \frac{1}{2}
	\label{eq:equacao2}
\end{equation}
Notice that the Born Rule maps state measurements to the concept of probability, i.e., the sum of the squares of the amplitudes of all possible states in the superposition is equal to 1:
\begin{equation}
	\abs{\alpha}^2 + \abs{\beta}^2 = 1
	\label{eq:equacao4}
\end{equation}
For the example in Listing \ref{alg:algoritmo-running-example1}, the qubits are measured in line 13 (\textit{measure\_all} method). Based on the Born Rule, $q_0$ will have a 50\% chance of collapsing to $\ket{0}$ and 50\% of collapsing to $\ket{1}$, as follows:
$$ \left \lvert \frac{1}{\sqrt{2}}\right \rvert  ^2 +  \left \lvert \frac{1}{\sqrt{2}} \right \rvert  ^2 = 1
$$
However, since the example circuit is composed of $q_0$ and $q_1$, the result is a combined state between these two qubits, which is represented by the tensor product ($\otimes$). 

Another interesting characteristic of QC is \textit{Entanglement}, which can be defined as a physical phenomenon in which multiple qubits are correlated with each other, such as the measurement of one of them automatically triggers correlated states of the others, even if these qubits are separated by great distances. 

In mathematical terms, entangled qubits represent superposition states that are not separable, i.e., cannot be factored into product states. 
In Listing \ref{alg:algoritmo-running-example1}, the entanglement is achieved with the CNOT (Conditional NOT) operator (line 12), which acts on two qubits, a control qubit (which is in superposition) and a target qubit, as follows: 
\begin{itemize}
\item if the control qubit is $\ket{0}$, no action is taken on the target qubit.
\item However, in case the control qubit is $\ket{1}$, the target qubit is switched.
\item the control qubit remains the same.
\end{itemize}

With the circuit defined, the next step consists of assembling it and submitting it to a back-end for execution. For Listing \ref{alg:algoritmo-running-example1}, the circuit is submitted on line 19 to the Aer simulator (defined in line 16), which consists of an ideal (noise-free) simulator running in the local computer \cite{QiskitAerSimulator}. 
Given the probabilistic nature of quantum programs, experiments usually need to be executed multiple times (parameter \textit{shots}, defined in line 19 in the run method), so the results can be checked against a certain expected probabilistic distribution. Once the circuit is executed, the results are collected in a variable called count, which contains a map of the returned values as well as their frequency.
Lastly, the function \textit{plot\_histogram} will print the returned results on (the x-axis) and their respective frequencies (on the y-axis).

Because of the differences in devices in terms of quantum architectures and hardware implementation details, there is a process called transpilation, which is responsible for rewriting a given input circuit to match the topology of that specific quantum device and optimizing the circuit \cite{qiskit2024}. In the running example, the transpilation is executed in line 17, by calling the method transpile with the circuit and the target execution environment (Aer simulator).

One key difference of QC applications when compared to their classical counterparts is the inability to clone quantum states. This is known as the \textit{No-cloning Theorem}. In classical computing, making multiple copies or inspecting the value of variables is a common task. The no-cloning theorem states that it is not possible to create a copy of an arbitrary quantum state \cite{stancil2022principles}. Thus, a read operation of an intermediate state in a qubit will make the quantum state collapse to a classical value, which will be the output state. Thus, the no-cloning theorem poses an important limitation to debugging quantum programs.

\begin{table}[h!]
\centering
\caption{Main Quantum Computing gates}
\renewcommand{\arraystretch}{1.5} 
\begin{tabular}{cc|cc}
\hline
\textbf{Gate name} (\textbf{Symbol}) & \textbf{Matrix} & \textbf{Gate name} (\textbf{Symbol}) & \textbf{Matrix}\\
\hline
\rule{0pt}{0.9cm} Hadamard (H) & $\frac{1}{\sqrt{2}}\begin{pmatrix} 1 & 1 \\ 1 & -1 \end{pmatrix}$ & \multirow{4}{*}[0.25cm]{CNOT (CNOT)} & \multirow{4}{*}[0.25cm]{$\begin{pmatrix} 1 & 0 & 0 & 0 \\ 0 & 1 & 0 & 0 \\ 0 & 0 & 0 & 1 \\ 0 & 0 & 1 & 0 \end{pmatrix}$}\\
\cline{1-2}
\rule{0pt}{0.9cm} Pauli-X (NOT) (X) & $\begin{pmatrix} 0 & 1 \\ 1 & 0 \end{pmatrix}$ \\
\hline
\rule{0pt}{0.9cm}  Pauli-Y (Y) & $\begin{pmatrix} 0 & -i \\ i & 0 \end{pmatrix}$  & \multirow{4}{*}[0.25cm]{Swap (SWAP)} & \multirow{4}{*}[0.25cm]{$\begin{pmatrix} 1 & 0 & 0 & 0 \\ 0 & 0 & 1 & 0 \\ 0 & 1 & 0 & 0 \\ 0 & 0 & 0 & 1 \end{pmatrix}$}\\
\cline{1-2}
\rule{0pt}{0.9cm} Pauli-Z (Z) & $\begin{pmatrix} 1 & 0 \\ 0 & -1 \end{pmatrix}$ \\
\hline
\rule{0pt}{2.5cm} \multirow{1}{*}[1cm]{Phase (S)} & \multirow{1}{*}[1.4cm]{$\begin{pmatrix} 1 & 0 \\ 0 & i \end{pmatrix}$} & \multirow{4}{*}[0.5cm]{Toffoli (CCNOT)} & \multirow{4}{*}[1.9cm]{$\begin{pmatrix} 1 & 0 & 0 & 0 & 0 & 0 & 0 & 0 \\
0 & 1 & 0 & 0 & 0 & 0 & 0 & 0 \\
0 & 0 & 1 & 0 & 0 & 0 & 0 & 0 \\
0 & 0 & 0 & 1 & 0 & 0 & 0 & 0 \\
0 & 0 & 0 & 0 & 1 & 0 & 0 & 0 \\
0 & 0 & 0 & 0 & 0 & 1 & 0 & 0 \\
0 & 0 & 0 & 0 & 0 & 0 & 0 & 1 \\
0 & 0 & 0 & 0 & 0 & 0 & 1 & 0 \end{pmatrix}$}\\
\cline{1-2}
\rule{0pt}{2.5cm} \multirow{1}{*}[1cm]{T Gate (T)} & \multirow{1}{*}[1.4cm]{$\begin{pmatrix} 1 & 0 \\ 0 & e^{i\pi/4} \end{pmatrix}$} \\
\hline
\end{tabular}

\label{tab:main_quantum_gates}
\end{table}

There are also gates such as $R_x$, $R_y$, and $R_z$ (Table \ref{tab:quantum_rotation_gates}) that are defined in terms of angles or rotations. These gates accept an angle as a parameter and rotate the target qubit around their respective axis by the specified angle. They are used in variational quantum algorithms, which are the basis of many Quantum Machine Learning (QML) algorithms. QML is currently an active research area in QC and has emerged as a dominant paradigm for circuit-based quantum programs in the current NISQ era \cite{simeone2022introduction}.

\begin{table}[h!]
\centering
\caption{Quantum rotation gates}
\renewcommand{\arraystretch}{1.5}
\begin{tabular}{c|c}
\hline
\textbf{Gate name} (\textbf{Symbol}) & \textbf{Matrix} \\
\hline
Rotation-X ($R_X(\theta)$) & 
$\begin{pmatrix} \cos(\theta/2) & -i\sin(\theta/2) \\ -i\sin(\theta/2) & \cos(\theta/2) \end{pmatrix}$ \\
\hline
Rotation-Y ($R_Y(\theta)$) & 
$\begin{pmatrix} \cos(\theta/2) & -\sin(\theta/2) \\ \sin(\theta/2) & \cos(\theta/2) \end{pmatrix}$ \\
\hline
Rotation-Z ($R_Z(\theta)$) & 
$\begin{pmatrix} e^{-i\theta/2} & 0 \\ 0 & e^{i\theta/2} \end{pmatrix}$ \\
\hline
\end{tabular}
\label{tab:quantum_rotation_gates}
\end{table}

In QML applications, there is typically a predefined parameterized quantum circuit (or variational circuit) with predefined architectures whose parameters are optimized by classical optimization algorithms. They are composed of different elements such as parameterized gates and entangling blocks. The classical data are coded into quantum states using different techniques and fed into the parameterized circuit. The output is read and sent to the classical optimizer. Some of the techniques for coding the classical data into quantum states are Basis Encoding, Amplitude Encoding, Time-Evolution Encoding, and Hamiltonian Encoding \cite{Schuld2021Machine}.
\begin{enumerate}
    \item \textit{Basis encoding}. It maps a classical value as an n-bit string to its computational basis state of an n-qubit system. For instance, the classical value  2 (0010) is mapped to the quantum state $\ket{2}$ ($\ket{0010}$).
    \item \textit{Amplitude encoding}. It associates classical data such as a real vector with quantum amplitudes. A normalized classical \(N\)-dimensional datapoint \( \mathbf{x} \) is represented by the amplitudes of a \( n \)-qubit quantum state \( |\psi_{\mathbf{x}}\rangle \) as \[|\psi_{\mathbf{x}}\rangle = \sum_{i=1}^{N} x_i |i\rangle.\] For example, let us say we want to encode the four-dimensional floating-point array \[\mathbf{x} = (1.0, 2.0, 3.0, 4.0)\] using amplitude embedding. The first step is to normalize it, i.e.,\[\mathbf{x}_{\text{norm}} = \frac{1}{\sqrt{30}} (1.0, 2.0, 3.0, 4.0)\]
    The corresponding amplitude encoding uses two qubits to represent \(\mathbf{x}_{\text{norm}}\) as
        \[
        |\psi_{\mathbf{x}_{\text{norm}}}\rangle = \frac{1}{\sqrt{30}} \left( 1.0 |00\rangle + 2.0 |01\rangle + 3.0 |10\rangle + 4.0 |11\rangle \right)
        \]
    \item \textit{Time-Evolution Encoding}. It maps a scalar \( x \in \mathbb{R} \) to the time parameter \( t \) in the unitary evolution of the Hamiltonian \( H \) defined as:
        \[
        U(x) = e^{-ixH}
        \]

        where \( U(x) \) is the unitary operator governing the evolution, \( x \) is the time parameter, and \( H \) is the Hamiltonian of the system. The exponential term \( e^{-ixH} \) describes the time evolution of the quantum state, with \( i \) as the imaginary unit, ensuring unitarity of the operation. 
    
        For instance, for the scalar value \(0.7\), when applying time-evolution encoding with an $R_Y(\theta)$ (Table~\ref{tab:quantum_rotation_gates}) gate and an initial state \(|0\rangle\), the resulting quantum state is given by:
            \[
            |\psi(0.7)\rangle = \cos\left(\frac{0.7}{2}\right)|0\rangle + \sin\left(\frac{0.7}{2}\right)|1\rangle
            \]
            \[
            |\psi(0.7)\rangle \approx 0.987|0\rangle + 0.173|1\rangle
            \]

    \item \textit{Hamiltonian Encoding.} It encodes matrices into the Hamiltonian of a time evolution by associating a Hamiltonian H with a square matrix A. For the cases in which A is not Hermitian ($A = A^\dagger$, i.e., A is equal to its conjugate transpose), the Hamiltonian matrix \( H \) can be represented as:
        \[
        H = \begin{pmatrix}
        0 & A \\
        A^\dagger & 0 
        \end{pmatrix}
        \]
        For example, if \( A \) is:
        \[
        A = \begin{pmatrix}
        1 & 2 \\
        3 & 4
        \end{pmatrix}
        \]
        Then, the conjugate transpose \( A^\dagger = A^T \) (since the matrix is real) will be:
        \[
        A^\dagger = \begin{pmatrix}
        1 & 3 \\
        2 & 4
        \end{pmatrix}
        \]
        Now, the Hamiltonian \( H \) becomes a 4x4 matrix:
        \[
        H = \begin{pmatrix}
        0 & A \\
        A^\dagger & 0
        \end{pmatrix}
        =
        \begin{pmatrix}
        0 & 0 & 1 & 2 \\
        0 & 0 & 3 & 4 \\
        1 & 3 & 0 & 0 \\
        2 & 4 & 0 & 0
        \end{pmatrix}
        \]

\end{enumerate}

These techniques implement the bridge between the classical world with classical data to the quantum world with quantum states. For the output, common methods to map measurement results to a class or label in a classification problem include measuring a single qubit, as proposed by \citeauthor{QuantumClassifiersMariaShuld} \cite{QuantumClassifiersMariaShuld}, or measuring all qubits and applying a post-processing function, such as calculating the parity \cite{Parity_Havl_ek_2019} or performing maximum-likelihood estimation on the measurement results, even for binary classification tasks \cite{Zhang_2022_Fast_Decay}. In cases where only a single qubit is measured, the state $\ket{0}$ typically represents one class, while $\ket{1}$ represents the other in a binary classification. When all qubits are measured, the mapping to the classes depends on the chosen function, which translates the combined measurement outcomes into distinct labels.

\section{Current testing and debugging approaches for quantum computing}
\label{sec:current-approaches}
This section provides an overview of key testing and debugging strategies for quantum programs (QPs). We first cover testing methods such as combinatorial testing, search-based testing, fuzz testing, mutation testing, and metamorphic testing. The concept of coverage is also discussed, as it can assume interesting forms in circuit-based programming model. Given the probabilistic nature of QPs, we explore the challenges of flaky tests and review current research on this issue. We then review techniques that leverage quantum properties like reversibility and unitarity, including equivalence checking and reversibility-based testing, as well as delta debugging combined with property-based testing. QuraTest is introduced as a tool for generating complex quantum inputs, such as superpositions or entanglements. In the second part, we discuss assertions for QPs, followed by a review of common bug patterns, taxonomies, and benchmarks. 

In terms of debugging techniques, we discuss the adaptation of classical methods for debugging QPs, such as backtracking and slicing. Quantum properties like superposition and no-cloning limit the development of these techniques. We present the Cirquo framework, which adapts classical slicing to quantum circuits by dividing them into smaller blocks. Coarse- and fine-grained methods are also covered, including the inspection of intermediate states in subroutines. Additionally, we introduce techniques such as probabilistic quantum cloning, used in tools like QDebug, to infer qubit states without collapsing them.

\subsection{Testing strategies}

In what follows, we describe the most promising testing techniques for QPs. Classical testing strategies developed for QC cover different aspects of a QP. For instance, Quito (QUantum InpuT Output coverage) \cite{Ali2021Assessing} is a framework to tackle the problem of test oracles for QPs, as well as coverage criteria for the program input, output, and input-output relations. The framework takes into account a Program Specification (PS) and uses statistical analysis of the test results to determine the criteria for passing the tests. Two Test Oracles are defined: Wrong Output Oracle (WOO) and Output Probability Oracle (OPO). 
Another classical approach used to tackle the dimensional problem of the input space is equivalence class partitioning, which is a functional testing technique that consists of dividing the input domain into different classes in which the software being tested has supposedly the same behavior \cite{tsui2022essentials}. 
For QPs, \citeauthor{TestingQPMultipleSubroutines} \cite{TestingQPMultipleSubroutines} have defined two equivalence partitioning criteria, namely:
\begin{enumerate}
    \item \textit{Classical-Superposition Partition (CSP)} -- each input variable (qubit) of each quantum state type is divided into classical state input and superposition state input.
    \item \textit{Classical-Superposition-Mixed Partition (CSMP)} -- each input variable of each quantum state type is divided into classical state input, superposition state input, and mixed state input.
\end{enumerate}
In terms of general classical testing approaches adapted for the testing of QPs, a few other examples in the literature are detailed in the following subsections. 

\subsubsection{Combinatorial testing} 
\leavevmode

The coverage criterion defined in Quito has scalability issues as the number of qubits increases. Thus, researchers have been working on optimization approaches to limit the input space. An example of such an approach is QuCAT (QUantum CombinAtorial Testing) \cite{Wang_2021}, a framework that applies Combinatorial Testing (CT) to generate tests for QP. The idea is that faulty points in the program can be reached through a particular combination of the input values of a given characteristic such as pair-wise or 3-wise.

\subsubsection{Search-based testing}
\leavevmode

A test generation tool for quantum programs, QuSBT (Search-based Testing of Quantum Programs) \cite{Wang_2022}, uses genetic algorithms to generate a test suite for quantum programs with the maximum number of failing tests. The input for the tool consists of the quantum program under test, a list of the input and output qubits, the total number of qubits, and a program specification (PS). The PS maps each input and output pair to a probability of occurrence. A statistical test (Pearson's chi-square) is used to check failures with a probabilistic nature, in which the user can also specify the significance level for the test.

\subsubsection{Fuzz testing}
\leavevmode

\citeauthor{Wang_2021_Fuzz} \cite{Wang_2021_Fuzz} investigated the use of Fuzz Testing \cite{fuzzingbook2024} in the generation of rare inputs for QPs to trigger sensitive branches and thus induce crashes or discover defects. The idea is to use a gray box testing approach to first identify the code measurement operations and the branches produced by these measurements. The procedure continues by producing input matrices that will maximize the probability of these sensitive branches being triggered, thus reaching the code with the possible defect. As measurement operations make qubits collapse to classical values, this approach is more of a hybrid quantum-classical technique, as everything after the measurement is purely classical and does not depend on any quantum characteristic. 

\subsubsection{Property-based testing}
\leavevmode

The probabilistic nature of QC programs makes it difficult to assert the value of certain quantum states, especially for test cases in which superposition plays a role. Property-based testing has been studied as an alternative to mitigate the non-deterministic nature of QPs \cite{PropertyBasedHonarvar}, as its main approach consists of generating tests based on general properties of the artifact being tested and not in concrete test cases. As with the classical property-based testing approach, the properties are described as pre-conditions as well as post-conditions and concrete program states become higher-level abstractions. The authors created a property specification language for testing Q$\#$ programs and a property-based testing method to generate, execute, and statistically assert the results of concrete test cases. To assert the test results, the authors define five types of assertions: Assert Probability, Assert Entanglement, Assert Equal, Assert Teleported, and Assert Transformed. These assertions rely on post-measurements and can generally be considered subsets of the more statistical and comprehensive checks in the work by \citeauthor{HuangStatisticalAssertions} \cite{HuangStatisticalAssertions}. However, \citeauthor{PropertyBasedHonarvar} \cite{PropertyBasedHonarvar} offer unique checks for quantum-specific protocols like teleportation, which is not explored in previous works on quantum assertions.

\subsubsection{Mutation testing}
\label{mutation_testing}
\leavevmode

Mutation testing plays two important roles in QPs: (1) mutation operators can be used to create faulty versions of QPs \cite{Ali2021Assessing}, thus mitigating the lack of quantum bug repositories and quantum benchmark programs, and (2) to assess the quality of test suites for QP. In terms of the assessment of the quality of test suites, \citeauthor{ArcainiMuskit} \cite{ArcainiMuskit} developed a mutation analysis tool for QPs called Muskit (MUtation testing for QisKIT) based on the Qiskit framework. Muskit has three components: Mutants Generator, Mutants Executor, and a Test Analyzer. The mutant generator component defines three mutation operators: Add Gate, Remove Gate, and Replace Gate. Similarly, \citeauthor{MutationTestingQiskitFortunato} \cite{MutationTestingRuiabreu1, MutationTestingRuiabreu2,MutationTestingQiskitFortunato} investigated the application of mutation testing in QPs written in Qiskit. The authors created a set of mutation operators to generate mutants based on qubit measurements and quantum gates. These operations were incorporated in a framework called QMutPy, which consists of an extension of the MutPy \cite{mutpy}, a Python-based tool for mutation testing. QMutPy extends the mutation operators already present in MutPy with five quantum operators: Quantum Gate Replacement (QGR), Quantum Gate Deletion (QGD), Quantum Gate Insertion (QGI), Quantum Measurement Insertion (QMI), and Quantum Measurement Deletion (QMD), which derive from the classical mutant operations.

\subsubsection{Metamorphic testing}
\leavevmode

Previous work \cite{MetamorphicTestingRuiAbreu} developed an approach to use metamorphic relations to test quantum programs. The authors use metamorphic rules, written as quantum functions and based on properties of the QP, to avoid or delay direct qubit measurement. These metamorphic relations are written as functions that can be executed directly in a quantum computer, consisting of what the authors call an ``oracle quantum program.'' Another approach developed by \citeauthor{PaltenghiMorphQ} \cite{PaltenghiMorphQ} is MorphQ, a metamorphic testing framework that aims to tackle two challenges related to testing QPs: (1) the lack of quantum programs available for testing; and (2) the oracle problem. MorphQ is equipped with an automatic quantum program generator, which uses both template-based and grammar-based code generation. The resulting programs do not crash during execution, as the generating strategies consider domain-specific constraints of quantum computing. To alleviate the oracle problem, MorphQ implements ten metamorphic transformations, such that the source and the follow-up programs have related outputs with equivalent behaviors.

\subsubsection{Flaky tests}
\leavevmode

Some works focus on the behavior of the quantum tests themselves with the study of flaky tests. These are tests that present non-deterministic behavior in which sometimes they pass and other times they fail. In this area, \citeauthor{FlakinessQPs} \cite{FlakinessQPs} manually analyzed 14 quantum software repositories and found out that 12 of them contained at least one flaky test. The authors identified eight categories of causes for flaky tests in these repositories, along with seven common fixes applied by programmers. The causes of flakiness fall into the following categories: randomness, software environment issues, multi-threading, floating-point operations, visualization, unhandled exceptions, network-related problems, and issues with Python data structures such as unordered collections. The common fixes include setting seeds in Pseudo-Random Number Generators (PRNGs), updating the development environment, using single-threaded routines, adjusting tolerance levels, adding exception handling, synchronizing network calls, and using keys for ordering in data structures rather than relying on insertion order. The causes of flakiness found by the authors are coherent with the non-deterministic behavior of QC as well as with the current status of the area, in which frameworks and programming languages are still under development and instabilities with different libraries and versions are expected.
Continuing their work in \cite{FlakinessQPs}, \citeauthor{FlakinessQPs2} \cite{FlakinessQPs2} propose a set of techniques to automate the detection of flakiness in bug reports. The authors suggest a machine learning based approach, in which a dataset with bug reports and labeled entities would be used to train a classification model. Each entity of the dataset would consist in a set of features extracted from the bug report and a classification (flaky or not flaky). As suggested by the authors, the model can be trained to detect flakiness in bug reports. A few of the challenges pointed out by the authors with this approach include data quality (as labels are extracted from text written by the issue reporter), insufficient data (as the field of QC is still developing), imbalance datasets, and dataset obsolescence.

\subsubsection{Coverage testing}
\leavevmode

\label{coverage}As defined by \citeauthor{Ammann_Offutt_2016}~\cite{Ammann_Offutt_2016}, a coverage criterion is a rule or collection of rules that impose test requirements on a test set. More formally, given a set of test requirements \(TR\) for a coverage criterion \(C\), a test set \(T\) satisfies \(C\) if and only if for every test requirement \(tr\) in \(TR\), at least one test \(t\) in \(T\) exists such that \(t\) satisfies $tr$. In terms of QPs, coverage has been explored in different areas such as input and output domains and, more recently, as branch coverage \cite{BranchCoverage}.
Quito \cite{Wang_2021_quito}, for instance, defines three coverage criteria based on the inputs and outputs of quantum programs:
\begin{itemize}
	\item \textit{Input Coverage} requires a test for each valid input defined in the PS.
	\item \textit{Output Coverage}  requires that each expected output, as defined by the PS, be observed at least once.
	\item \textit{Input-Output Coverage} requires a test for each input-output pair. Similarly to the output criterion, each input/output pair is observed at least once.
\end{itemize}
The effectiveness of each criterion was evaluated using mutation analysis and the failing and passing of the test suites were evaluated using a one-sample Wilcoxon signed ranked test.

\citeauthor{BranchCoverage} \cite{BranchCoverage} define the concept of Gate Branch Coverage (GBC) for QPs. The authors propose a metric for calculating the coverage of branches generated by controlled gates such as the CNOT. The metric is defined as follows:
$$
\text{GBC}(G) = \left(\frac{\sum_{g \in G} e(g)}{\sum_{g \in G} \left(1 + c(g)\right)}\right) \times 100\%
$$
where G is the set of controlled-type gates of the program (CNOT, CCNOT, CHadamard, etc.), e(g) is the number of exercised branches of a given gate g, and c(g) is the number of control qubits of a given gate g.  

The concept of cyclomatic complexity for QPs was introduced by \citeauthor{FormalizationCoverageCriteria}~\cite{FormalizationCoverageCriteria}. This study outlines three coverage criteria for structural testing:
\begin{enumerate}
    \item \textit{Single Qubit Gate Coverage Criteria}. This metric is calculated based on the ratio between the number of times a certain gate is executed by the tests and the total number of times that gate appears in the circuit under test. For instance, let the X-gate coverage of test suite T for a QP be denoted as $X_c/X_T$. Here, $X_c$ represents the number of times the gate X was executed by any test in the circuit, and $X_T$ is the number of times X appears in the circuit.
    
    Note that in the circuit-based model every single-qubit gate will be executed by the quantum computer whenever the circuit is submitted. As a result, all single-qubit gates are applied, even if they do not always have a measurable effect on the qubits. Therefore, under the Single Qubit Gate Coverage Criteria defined above, the coverage for single-qubit gates in the circuit under test will always be 100\%, given that there is at least one test that submits the circuit to the quantum computer.
    \item \textit{Two Qubit Gate Coverage Criteria}. Two qubit gates such as CX, CY and CZ (controlled versions of the gates presented in Table~\ref{tab:main_quantum_gates}) can trigger three different paths: the case where the control qubit is set to $\ket{0}$, to $\ket{1}$ and when it is in superposition. 
    For any given two-qubit gate, the coverage criterion is defined as:

        $$ 
            CGate_{cr} = \frac{(|T| + |F| + |S|)}{3 \times |K|}
        $$ 
        where
        \begin{itemize}
            \item |T|: The total number of times the given gate appears in the circuit  with control qubit set to $\ket{1}$ by test cases from the test suite.
            \item |F|: The total number of times the given gate appears in the circuit  with control qubit set to $\ket{0}$ by test cases from the test suite.
            \item |S|: The total number of times the given gate appears in the circuit  with control qubit set to superposition by test cases from the test suite.
            \item |K|: The total number of times the given gate appears in the circuit.
        \end{itemize}
    
    \item \textit{Three Qubit Gate Coverage Criteria.} The two gates considered in the coverage criteria are the CCNOT (Toffoli) and the CSWAP gates. The Toffoli Gate Coverage criterion defines a measure in which all Toffoli gates are executed and (i) the target qubit remains the same, and (ii) X-gate is applied to all Toffoli paths covered by test suites. Two test cases are needed to consider the superposition of both control qubits.
    $$ 
    TF_{cr} = \frac{|F| + |T| + |FQ| + |SQ|}{4 \times |TT|}
    $$ 

    where
    \begin{itemize}
    \item |F| - number of qubits flipped by the Toffoli gates;
    \item |T| - number of target qubits remaining the same; 
    \item |FQ| - Toffoli gates executed by the first control qubit in superposition;
    \item |SQ| = number of Toffoli gates executed by the second control qubit in superposition;
    \item |TT| - total number of Tofolli gates in the circuit.
    \end{itemize}
    Similarly, the Fredkin (CSWAP) coverage criterion is defined based on the number of test cases in the test suite that covers the scenarios in which the control qubit is either $\ket{1}$ (swapping occurs), $\ket{0}$ (no swapping occurs) or in a superposition of these states. 
\end{enumerate}
The authors also define a criterion for multiple controlled qubit gates which is based on the premise that test cases will execute all combinations of various control qubits.

The coverage criterion defined by \citeauthor{BranchCoverage} \cite{BranchCoverage} is similar to the Two Qubit Gate Coverage Criteria defined by \citeauthor{FormalizationCoverageCriteria}~\cite{FormalizationCoverageCriteria}. While the first criterion focuses on the classical branches that are generated by control qubits in controlled gates, it however does not consider superposition, which can be a possible state of the control qubit(s), thus triggering multiple branches simultaneously. An example of such a scenario can be illustrated by the Quantum Walks algorithm \cite{Kempe_2003}. On the other hand, $CGate_{cr}$ measures coverage by including all three possible states of the control qubit ($|0\rangle$, $|1\rangle$, and superposition).

\subsubsection{Other quantum-based testing approaches}
\leavevmode

\label{other_testing_approaches}
The characteristics of QC described in Section \ref{sec:quantum-computing}, notably superposition and the impossibility of cloning quantum states, pose interesting challenges for researchers in Quantum Software Testing. However, other particularities of quantum systems have been leveraged in the development of novel testing approaches. Reversibility and the unitary nature of quantum operations are cornerstone concepts in specific testing techniques, including: (i) partial equivalence checking \cite{PartialEquivalenceChecking}, (ii) fast equivalence checking for quantum circuits \cite{FastEquivalenceCheckingQP}, (iii) fault testing for reversible circuits \cite{FaultTestingReversibleCircuits}, (iv) methods for k-CNOT gates (multiple-control Toffoli gates) \cite{ADesignForTestability}, and (v) design techniques for gates such as Toffoli, Fredkin, and mixed Toffoli-Fredkin that aim to improve circuit testability \cite{TestableDesignsofToffoli}. 

In terms of equivalence checking, \citeauthor{PeixunJianjunEquivalence} \cite{PeixunJianjunEquivalence} propose three algorithms to address key challenges: (1) Equivalence Checking, which checks whether two quantum programs $ \mathcal{P} $ and $ \mathcal{P}'$ are equivalent, meaning that, both programs produce the same quantum output for the same quantum input; (2) Identity Checking, which verifies whether a quantum program $ \mathcal{P} $ represents an identity transform; and (3) Unitarity Checking, which ascertains whether a quantum program $ \mathcal{P} $ represents a unitary transform. These differ from circuit optimization techniques, for instance, in Qiskit, as the authors work in the context of black-box testing, where the internal structure of the program is not directly considered. 

For equivalence checking, the authors developed a method based on the Swap Test, which compares the outputs of two QPs with input states based on the Pauli states (the eigenstates of the Pauli matrices) in order to determine whether or not two programs are equivalent. The technique compares only a subset of the outputs and does not explore all possible input states. The rationale behind this choice is that Pauli states form a complete basis for quantum operations, so if two quantum programs behave equivalently on a random selection of these states, they are likely to behave equivalently on most other input states. If, after running the test on this subset, the outputs are similar (within a defined tolerance), the algorithm returns that the programs are equivalent. If discrepancies are found, the programs are marked as non-equivalent. 

For the Identity Checking, the idea is to check whether a certain QP is equivalent to an identity operation, i.e., it leaves all input states unchanged. The algorithm begins by applying a randomly chosen Pauli input state to the program. In case the program is the identity operation, the output state should be exactly the same as the input state (the Pauli input state). To verify this, after the program runs, an inverse transform of the initial random input state is applied. If the program has performed the identity operation correctly, this inverse transform will always return the system to the initial basis state, usually $\ket{0}^{\otimes n}$. A measurement is then made, and if any result other than $\ket{0}^{\otimes n}$ is returned, the program is not the identity. The algorithm tests multiple random input states to confirm whether the identity operation holds across a range of possible inputs.

For the Unitarity checking, the authors use the fact that unitary operations preserve the inner product between states and that they map orthogonal input states to orthogonal output states,  preserving the purity of pure states. The check involves two tests: first, verifying that pairs of orthogonal computational basis states remain orthogonal after being processed by the program, and second, confirming that superposition states, such as $\frac{1}{\sqrt{2}}(\ket{m} + \ket{n})$ and $\frac{1}{\sqrt{2}}(\ket{m} - \ket{n})$, maintain their orthogonality after the program's action. The Swap Test is used to measure the overlap of output states, and the test returns a fail if the outputs exhibit non-zero overlap, indicating that the transformation is not unitary. By sampling a subset of input states, the algorithm verifies that the program is equivalent to a unitary operation.

Other approaches such as Qraft \cite{Qraft} are fully based on the reversal of the circuit being tested to predict the program output. The idea of Qraft is to reverse the quantum circuit and execute the full forward as well as its reversed version to deduce the correct program output for all the quantum states of the original (forward) circuit. Similarly, Qdiff \cite{Qdiff}, a differential testing framework for testing QPs, takes advantage of reversibility and unitary operators to generate logically equivalent variants of the QPs being tested. Finally, Miranskyy's \cite{MiranskyyDynamicTesting} approach considers using QCs to speed up dynamic tests of classical programs. Although the focus of his work is not the test of QPs \textit{per se}, the author shows that, in some cases, it is possible to translate a classical program to a quantum program and then take advantage of QC computational power. 
Other strategies combine different techniques for testing and debugging QPs. For instance, \citeauthor{Pontolillo2024Delta} \cite{Pontolillo2024Delta} integrate delta-debugging and property-based testing for automated debugging QPs. The authors use three quantum algorithms (Quantum Fourier Transform, Quantum Teleportation, and Quantum Phase Estimation) to evaluate the developed technique. Their approach consists of using the correct implementation of each algorithm and inserting defects along with semantic-preserving changes (such as two identical operators in a row, resulting in an Identity operator). Then the process aims to correctly identify the defect inserted in the output while excluding the changes that had no effect on the result (semantic-preserving changes).

Most of the techniques discussed in this section consider only basic states $(\ket{0}$ and $\ket{1}$) as the inputs for the QPs) being tested. However, quantum states can also exist in superposition, become entangled, and have characteristics such as phase. While generating test cases using only the base states can be effective in certain applications, it may not be sufficient to cover the entire input space. To fill this gap, \citeauthor{QuraTest} \cite{QuraTest} created QuraTest, an automated testing framework for generating inputs for QPs that use three test input generators (UCNOT, IQFT and Random) to create complex input states in terms of magnitude, phase, and entanglement. The results show that QuraTest and its test generation techniques can create more diverse test cases that increase output coverage and mutation score. Another technique to handle the challenges with the input states for test cases is QOPS \cite{muqeet2024quantumprogramtestingcommuting}. Tailored to handle input states in superposition, which are used by optimization and search algorithms, QOPS employs Pauli strings (tensor products of Pauli operators) to define test cases for quantum programs. Unlike traditional methods that require explicit program inputs, QOPS focuses on the measurement operation rather than input initialization, making it suitable for quantum algorithms that rely on superposition, such as Grover’s and QAOA. The approach leverages the commuting property of Pauli strings, which allows multiple observables to be measured simultaneously, reducing the need for exhaustive program specifications. Additionally, QOPS integrates with industrial error mitigation techniques, such as IBM's Estimator API, allowing it to test quantum programs on real noisy quantum computers. 

Another example of a combination of techniques for testing QPs is the work by \citeauthor{MutationSearchSimula} \cite{MutationSearchSimula}, which integrates Mutation Testing with a multi-objective search-based approach (NSGA-II) for test case generation. The framework developed by the authors is designed to generate a minimal number of test cases to maximize the number of mutants killed in a QP, as well as to identify equivalent programs that cannot be killed, allowing them to be filtered out of the analysis. Given that the access to real QCs for test execution is still limited, these strategies are valuable as they reduce the testing effort compared to the baseline approach (random search).

The techniques based on multiple executions of a QP (multiple samples) described in this section perform well for small QPs, but face scalability issues for larger QPs, such as those with 60 qubits, where these techniques may not be able  to handle the vast search space ($2^{60}$). This issue is tackled in the work by \citeauthor{oldfield2024fasterbetterquantumsoftware} \cite{oldfield2024fasterbetterquantumsoftware}. As QPs scale to larger qubit counts, the number of possible outcomes grows exponentially, leading to the need for an increasing number of measurements to validate the program's correctness. The authors also emphasize that current testing techniques for QPs  do not cover quantum-specific bugs, such as phase flip faults, which cannot be detected when using standard computational basis measurements and only become detectable through projective measurements in the Hadamard basis. To address these issues, they propose a Greedy reduction-based approach that reduces the complexity of a PS, which enables more efficient sampling. By lowering the number of basis states in the PS, the method optimizes testing without sacrificing the ability to detect quantum-specific faults.

There are also works focusing on reducing the impact the noise has in the testing techniques. For instance, \citeauthor{MitigatingNoiseMuqeet} \cite{MitigatingNoiseMuqeet} developed QOIN(Quantum sOftware testIng Noise-aware), a machine-learning-based tool to mitigate the impact of noise on QP testing. The framework consists in training a neural network to learn noise patterns specific to quantum computers and then applying transfer learning to adapt to individual quantum circuits. The authors highlight the fact that the noise patterns are not only specific to quantum computers but also to the quantum circuits being executed. The filtered output produced by the framework allows for more accurate test case assessments, even on noisy quantum computers. The authors evaluate QOIN using noise models from IBM, Google, and Rigetti, demonstrating its ability to reduce the noise effect by over 80\% across most models. However, QOIN has a few limitations that are important to consider such as poor performance in some of the quantum backends used in the experiments as well as in circuits with a high number of phase gates.

\subsection{Assertions} 
\label{sec:Assertions}

Assertions in QPs are an important area of study due to the unique constraints posed by quantum computing (Section \ref{sec:quantum-computing}). These include non-deterministic outcomes causing the oracle problem and flakiness in tests \cite{FlakinessQPs}, and the no-cloning theorem. The probabilistic behavior also studied in the testing of classical programs is particularly relevant due to the uncertain nature of quantum states. Thus, for the programmer creating tests for a quantum routine, it is considerably difficult to define an oracle upon which the assertions can be based.
The inability to directly read quantum states without collapsing them to classical values makes it impossible to use assertions in the middle of a QP. Consequently, the approaches that emerged to tackle the challenges associated with assertions in QPs are:
\begin{enumerate}
    \item \textit{Measurement-based Assertions}.  These assertions measure the qubit under test at a certain point during the program execution. As such, the state of the qubit collapses and the subsequent execution of the program is impacted. These assertions assume that the program needs to be executed multiple times to perform statistical tests and determine the possible state of the qubit given a certain significance level. Examples of measurement-based assertions are Statistical Assertions \cite{HuangStatisticalAssertions}, Assertions using Swap-Tests \cite{SwapTestAssertions}, and Projection-based Assertions \cite{ZhouProjectionBasedAssertions}.
    
    \item \textit{Quantum-based Assertions}. These assertions are usually called runtime or dynamic assertions and do not rely on measuring the assessed qubit. As the qubit is not measured, the program state is not impacted as the quantum state does not collapse to a classical value. The main types of quantum-based assertions are Runtime (Dynamic) Assertion checking \cite{LiuRuntimeAssertions}, Assertions for Memberships / Approximate Assertions, and Swap-based assertions \cite{SystematicApproachesAssertionsLiu}.

    \item Other types of assertions found in the literature explore particular characteristics of the circuit being tested. They are: Assertions for Symmetry States \cite{ExploitingQuantumAssertions},  Nondestructive Discrimination (NDD) assertions \cite{LiuRuntimeAssertions}, and Invariant and Inductive assertions \cite{YingInvariants}. 
    
\end{enumerate}

The remaining of this section presents each assertion type defined above with their respective sub-types.

\subsubsection{Measurement-based assertions} \leavevmode

Three types of measurement-based assertions are discussed below, namely,  Statistical-based,  Swap-test, and Projective-based assertions.

\textit{Statistical Assertions}. These are the most general types of measurement-based assertions. As defined by \citeauthor{HuangStatisticalAssertions} \cite{HuangStatisticalAssertions}, statistical assertions consist of analyzing the results (measurements) of multiple executions of a QP to estimate the probable state of a qubit being analyzed. Since a measurement makes the quantum states collapse to a classical value, such assertions terminate the quantum program. Continuing the program after measurement impacts subsequent calculations, as the quantum operators expect quantum states and not classical bits. This way, an assertion is equivalent to a vertical slice of a QP, similar to what has been recently proposed by \citeauthor{MetwalliToolDebugging}~\cite{MetwalliToolDebugging}. The authors defined three types of assertions for QPs:
\textit{Classical assertions} are used to verify initial qubit values or check intermediate classical states. These assertions check if the result of a quantum state measurement is equal to a classical value (an integer). 
\textit{Superposition assertions}  check if quantum variables are in a certain superposition state, i.e., the result of multiple measurements of the given state follows a probabilistic distribution.
\textit{Entanglement assertions}  check if two quantum particles are entangled.  

    \textit{Assertions using Swap-Tests}. This type of assertion uses a swap-test to assert whether or not two qubits are equal to each other. The swap-test circuit is presented in Figure \ref{fig:swap_test}. As described by \citeauthor{SwapTestAssertions}~\cite{SwapTestAssertions}, the swap-test uses three qubits: (1) a control qubit $\ket{\xi}$, initialized in the circuit with $\ket{0}$; and (2) the two qubits to be asserted for equality, here represented by $\ket{\psi} = \ket{a}$ and $\ket{\phi} = \ket{b}$.
    The algorithm has four steps:
    \begin{enumerate}
    \item A Hadamard gate is applied to $\ket{\xi}$, creating a superposition.
    \item A controlled SWAP gate is applied to swap $\ket{\psi}$ and $\ket{\phi}$ if $\ket{\xi} = \ket{1}$.
    \item Another Hadamard gate is applied to $\ket{\xi}$, and its value is measured.
    \item The probability of measuring 0, $Pr(\ket{0}) = \frac{1}{2} + \frac{1}{2} \left|\bra{b}\ket{a}\right|^2$, indicates how close the two qubits are to being equal.
    \end{enumerate}
    Based on the final equation, one can observe that if $\ket{b}=\ket{a}$, then $Pr(\ket{0})=1$, as $\bra{a}\ket{a}=1$ and $\frac{1}{2} + \frac{1}{2} \left|1\right|^ 2 = 1$.
    On the other hand, if $\ket{b}$ and $\ket{a}$ are orthogonal, the product $\bra{a}\ket{b}$ results in 0. Thus, $Pr(\ket{0})=\frac{1}{2}$.
    The process consists in running the swap-test several times. If the result of every execution is $\ket{0}$ (no assertion error), then $\ket{b}=\ket{a}$. Otherwise, if there is at least one result as $\ket{1}$, then $\ket{b}\neq\ket{a}$.   
    The multiple executions are needed because, when the first qubit (ancilla)  in the swap test is $\ket{0}$, it can be that the qubit under test and the desired state are orthogonal. This way, the ancilla qubit still has a 50\% probability of being $\ket{0}$ \cite{SystematicApproachesAssertionsLiu}.

\begin{figure}[ht]
    \begin{center}
        \begin{quantikz}[wire types={q,q,q,c}]
            \lstick{$0$} & \gate{H} & \ctrl{1} & \gate{H} & \meter{} \vcwdouble{1-5}{4-5}{0} & \\
            \lstick{$\ket{\phi}$} &          & \swap{1}      &   & & \\  
            \lstick{$\ket{\psi}$}     &       &    \swap{0}     &       &  & \\
            \lstick{c}     & \qwbundle{1}      &          &       &  & 
        \end{quantikz}
    \end{center}
    \caption{Circuit to assert whether a qubit is in superposition or not. Source: \cite{SwapTestAssertions}}
     \label{fig:swap_test}
\end{figure}
    
    \textit{Projection-based Assertions}.
    \label{subsec:ProjectionBasedAssertions}
    Projection-based assertions rely on measurement for assessing the state of a certain qubit of interest \cite{ZhouProjectionBasedAssertions}. This assertion type is based on projections, which are similar to closed subspaces of a state space. Similar to the assertions defined by
    \citeauthor{HuangStatisticalAssertions} \cite{HuangStatisticalAssertions}, projection-based assertions use statistical tests based on multiple executions of a function to assert its correctness. However, the advantage of the projection-based assertion is that they rely on a smaller number of projective measurements to verify predicates directly, resulting in a more efficient process compared to purely statistical assertions.

\subsubsection{Quantum-based assertions} \leavevmode
\label{subsec:QuantumBasedAssertions}

Quantum-based assertions act directly on the qubits without performing any measurements. Thus, qubit states do not collapse and the program execution is not impacted. 

\textit{Runtime (Dynamic) Assertion checking}.
\label{subsec:RuntimeDynamicAssertions}
Inspired by quantum error correction, which uses ancilla qubits to encode quantum information and correct possible errors in qubits, \citeauthor{LiuRuntimeAssertions} \cite{LiuRuntimeAssertions} used ancilla qubits to indirectly verify the state of qubits under investigation. The idea is to measure ancilla qubits and infer the state of the actual qubits being analyzed.

The assertion types defined by the authors are consistent to those proposed by \citeauthor{HuangStatisticalAssertions} \cite{HuangStatisticalAssertions}, namely, classical, superposition and entanglement assertions. However, unlike statistical assertions, runtime assertions with ancilla qubits do not impact the state of the qubits being measured. For each assertion type, the authors also define the respective circuit that can be used to implement each assertion in execution time.

The circuit for the dynamic assertion of classical values is illustrated in Figure \ref{fig:circuit1-classical}. The idea is to check if a qubit is correctly initialized (for instance, in the $\ket{0}$ state) or to assert the base values in intermediate calculations. The circuit consists of the state of interest, $\ket{\psi}$, and an ancilla qubit, which is used to avoid measuring $\ket{\psi}$ directly. Both qubits are initialized to $\ket{0}$.

If $\ket{\psi}$ is correctly initialized and remains in the $\ket{0}$ state, then the CNOT gate will not invert the state of the ancilla qubit, since the control qubit ($\ket{\psi}$) is in the $\ket{0}$ state. Consequently, the ancilla qubit will remain in the $\ket{0}$ state. However, if $\ket{\psi}$ is in the $\ket{1}$ state, based on the logic of the CNOT gate, the ancilla qubit will be flipped to the $\ket{1}$ state.

\begin{figure}[ht]
    \begin{center}
        \begin{quantikz}[wire types={q,q,c}]
            \lstick{$\ket{\psi}$} & \ctrl{1} &  & \\
            \lstick{ancilla} &   \targ{}  & \meter{} \vcwdouble{2-3}{3-3}{1}     &   \\  
            \lstick{c}     & \qwbundle{2}  &  & 
        \end{quantikz}
    \end{center}
   	  \caption{Circuit to assert the classical value of a qubit. Source: \cite{LiuRuntimeAssertions}}
	\label{fig:circuit1-classical}
\end{figure}

Although this assertion is intended to be applied to intermediate classical states, it is important to note that applying a CNOT gate to a qubit in superposition will entangle both qubits. In such a case, measuring the ancilla qubit, as proposed by this type of assertion, will cause both qubits to collapse to classical values, thereby affecting the subsequent execution of the circuit and invalidating the use of the assertion.

The circuit for the dynamic assertion of entanglement is presented in Figure \ref{fig:circuit2-entanglement}. The purpose of this circuit is to assert whether the qubits $\ket{\phi}$ and $\ket{\psi}$ are in the entangled state $a\ket{00} + b\ket{11}$ or $a\ket{01} + b\ket{10}$. The circuit checks the parity of the two qubits being asserted, using an ancilla qubit to avoid direct measurements. The ancilla qubit is initialized to $\ket{0}$ when asserting the entangled state $a\ket{00} + b\ket{11}$, or to $\ket{1}$ when asserting the state $a\ket{01} + b\ket{10}$

\begin{figure}[ht]
    \begin{center}
        \begin{quantikz}[wire types={q,q,q,c}]
            \lstick{$\ket{\psi}$} & \ctrl{2} &  &  &  \\
            \lstick{$\ket{\phi}$} & &  \ctrl{1} &  &  \\
            \lstick{ancilla}      & \targ{}  &   \targ{}  & \meter{} \vcwdouble{3-4}{4-4}{2}  &   \\  
            \lstick{c}     & \qwbundle{3}  &  &  & 
        \end{quantikz}
    \end{center}
      \caption{Circuit to assert the entanglement of two qubits. Source: \cite{LiuRuntimeAssertions}}
        \label{fig:circuit2-entanglement}
\end{figure}

A critical observation concerning the circuit for asserting entanglement (Figure \ref{fig:circuit2-entanglement}) is its limited ability to determine the presence of entanglement more broadly, specifically whether two qubits are entangled or not. Instead, the circuit is designed to detect entanglement in two or more qubits that exist in either of two specific states: $ \alpha \ket{00} + \beta \ket{11}$ or $\alpha\ket{01} + \beta\ket{10}$. For instance, let the ancilla qubit to be initialized in the state $\ket{0}$ to assess the entangled state $ \alpha \ket{00} + \beta \ket{11}$ between $\ket{\psi}$ and $\ket{\phi}$. If no entanglement exists between $\ket{\psi}$ and $\ket{\phi}$, the four possible scenarios are:

\begin{enumerate}
    \item $\ket{\psi}$ is in the state $\ket{0}$ and $\ket{\phi}$ is in the state $\ket{0}$. Neither the first CNOT gate nor the second CNOT gate change the state of the $q[2]$ (the ancilla qubit), as both $\ket{\psi}$ and $\ket{\phi}$ are in the $\ket{0}$ state. Thus, the circuit will return 0, meaning that both qubits are entangled in the state $ \alpha \ket{00} + \beta \ket{11}$, when, in fact, they are not entangled at all.
     
    \item $\ket{\psi}$ is in the state $\ket{0}$ and $\ket{\phi}$ is in the state $\ket{1}$. The first CNOT gate does not change the state of $q[2]$, as the control qubit $\ket{\phi}$ is in the $\ket{0}$ state. However, the second CNOT gate will change the state of $q[2]$ from $\ket{0}$ to $\ket{1}$, as the control qubit $\ket{\phi}$ is in the state $\ket{1}$. In this case, the assertion circuit returns 1, indicating that there is an assertion error.
    
    \item $\ket{\psi}$ is in the state $\ket{1}$ and $\ket{\phi}$ is in the state $\ket{0}$. Similarly to the previous case, the first CNOT gate will switch the ancilla qubit from $\ket{0}$ to $\ket{1}$, while the second CNOT gate will not have an effect. Thus, the assertion circuit returns 1, indicating that there is an assertion error for this combination as well.
    
    \item $\ket{\psi}$ is in the state $\ket{1}$ and $\ket{\phi}$ is in the state $\ket{1}$. As the ancilla qubit is initialized in the state $\ket{0}$, the first CNOT gate will switch it to $\ket{1}$, while the second CNOT gate will switch it back to $\ket{0}$. That indicates that the qubits are entangled in the  $\alpha \ket{00} + \beta \ket{11}$, when they are not entangled, as initially assumed.
\end{enumerate}

The circuit assumes prior knowledge of entanglement between the qubits and aims to determine the specific state of entanglement. In this context, the entanglement of formation metric is utilized to establish the entanglement of two qubits.

The dynamic assertion for superposition is presented in Figure \ref{fig:circuit3-superposition}.

\begin{figure}[ht]
    \begin{center}
        \begin{quantikz}[wire types={q,q,c}]
            \lstick{$\ket{\psi}$} & \ctrl{1} & \gate{H} & \ctrl{1} &  & \\
            \lstick{ancilla} &   \targ{}  & \gate{H}  & \targ{} & \meter{} \vcwdouble{2-5}{3-5}{1} &  \\  
            \lstick{c}     & \qwbundle{2}  &  & & & 
        \end{quantikz}
    \end{center}
    \caption{Circuit to assert whether a qubit is in superposition or not. Source: \cite{LiuRuntimeAssertions}}
    \label{fig:circuit3-superposition}
\end{figure}

This assertion circuit aims to determine whether the qubit under test is in a superposition state by entangling it with an ancilla qubit and measuring the ancilla. The circuit uses a combination of CNOT and Hadamard gates to create and manipulate the entanglement. If the qubit is in superposition, the measurement of the ancilla has a deterministic outcome, indicating whether the qubit is in the $ \ket{+}$ or $\ket{-}$ superposition. On the other hand, if the qubit is in a classical state, the measurement result of the ancilla will be probabilistic, reflecting the mixed nature of the entanglement. 
This assertion impacts the qubit being measured as described above. If the state being measured is in superposition, it will collapse to a classical value.

\textit{Assertions for Memberships/Approximate Assertions}. 
The work in \citeauthor{LiuRuntimeAssertions} \cite{LiuRuntimeAssertions} covers assertions for an even number of entangled states or an odd number of single entangled states, thus lacking support for other types of entanglement states such as the GHZ state (
$\ket{GHZ} = \frac{1}{\sqrt{2}}\ket{000} + \frac{1}{\sqrt{2}}\ket{111}$) \cite{Greenberger1989}.
Besides, the assertion types defined in their work also require that the developer knows beforehand the precise state information being asserted.
To circumvent these limitations, \citeauthor{SystematicApproachesAssertionsLiu} \cite{SystematicApproachesAssertionsLiu} extend the ideas in \cite{LiuRuntimeAssertions} with the definition of precise and approximate assertions. Precise assertions can check specific quantum states, while approximate assertions can check whether the qubits being asserted are part of a certain set (${\ket{\psi}, \ket{\theta}, ...}$) of expected states. Precise assertions are further divided into assertions for pure and mixed states.

\textit{Swap-based assertions}.
Swap-based assertions are similar to swap-test assertions in the sense that both use the swap gate to avoid direct measurement of the qubits under test. Likewise, both assertion types are used to check the equality of two qubits, i.e., $assertEquals(\ket{\psi},\ket{\phi})$ \cite{SystematicApproachesAssertionsLiu}. The main idea is that any n-qubit pure state can be generated by transforming the ground state $\ket{0}^{\otimes n}$ by the application of a certain unitary transformation U, i.e., a pure state in a quantum program can be described as a sequence of operations $V_n$ over $\ket{0}^\otimes$, resulting in a final state $\ket{\psi}$, i.e, $\ket{\psi}= V_1...V_n \ket{0} ^ {\otimes n}$. This way, to assert that $\ket{\psi}$ is equal to the target state $\ket{\phi}=U\ket{0}^{\otimes{n}}$, the inverse gate $U^{-1}$ needs to be applied to bring the pure state $\ket{\psi}$ back to the n-qubit state $\ket{0}^{\otimes n}$.
The algorithm requires a measurement step to assess that all the qubits are in fact in the $\ket{0}$ state. When that is not the case, the pure state $\ket{\psi}$ was not in the state $\ket{\phi}$ before the inverse transformation, raising an assertion error \cite{SystematicApproachesAssertionsLiu}.
By the time this work was published, modern quantum programming languages did not support operations after measurement, making it impossible to proceed with the program after measurement. At the time, the authors introduced SWAP gates and ancilla qubits to avoid direct measurement of the qubits being tested (only the ancilla qubits would be measured). The circuit for this assertion is presented in Figure \ref{fig:swap-based-assertions}.

\begin{figure}[ht]
    \begin{center}
        \begin{quantikz}[wire types={q,q,q,q}]
            \lstick[2]{$\ket{0}^{{\otimes}{n}}$} &  \gate[2]{V_{1}}\gategroup[2,steps=3,style={dashed,rounded
corners,inner xsep=2pt},background,label style={label
position=above,anchor=north,yshift=0.3cm}]{Quantum program} & \ \ldots\  & \gate[2]{V_{n}} & \rstick[2]{$\ket{\psi}$} & & \gate[2]{U^{-1}}\gategroup[5,steps=5,style={dashed,rounded
corners,inner xsep=2pt},background,label style={label
position=above,anchor=north,yshift=0.3cm}]{Assertion circuit} &  &  & \swap{3} & & & \rstick[2]{$\ket{\phi}$} \\
             &     & \ \ldots\  &  &  &  &  &  & &  & \swap{3} & &   \\  
            \ \ldots\ \\
            \lstick[2]{$\ket{0}^{{\otimes}{n}}$} &   &  & & &  & \gate[2]{U} & \rstick[2]{$\ket{\phi}$} &  & \swap{0} &  &  &  \meter{} \\
             &   &  &  & &  &  &  &  &  & \swap{0} & & \meter{}    
        \end{quantikz}
    \end{center}
  \caption{Swap-based assertions circuits. Source: \cite{SystematicApproachesAssertionsLiu}}
  \label{fig:swap-based-assertions}
\end{figure}

\subsubsection{Other assertion types}.
Based on error-mitigating techniques, \citeauthor{ExploitingQuantumAssertions} \cite{ExploitingQuantumAssertions} define the concept of \textit{Assertions for Symmetry States}. This assertion type consists of leveraging the specific structure of the circuit being tested to check if a certain state is symmetric regarding a given operation, i.e., the evolution of that state triggered by the operation is restricted to the corresponding eigenspace. 

\textit{Nondestructive Discrimination} (NDD) \textit{assertions} \cite{SystematicApproachesAssertionsLiu} are based on the NDD protocol, which is used in secure quantum communication. This assertion type can be applied to discriminate entangled states such as Bell states in quantum information processing \cite{LiuRuntimeAssertions}. In their work, \citeauthor{SystematicApproachesAssertionsLiu} \cite{SystematicApproachesAssertionsLiu} adapt the NDD protocols to assert certain quantum states without measuring the qubits under test, as described in Section \ref{subsec:RuntimeDynamicAssertions}. The authors developed these ideas further to consider assertions for pure and mixed states.

\citeauthor{YingInvariants} \cite{YingInvariants} define the concept of \textit{invariant and inductive assertions} to verify the partial correctness of quantum programs, which can be of two types as defined by the authors: additive and multiplicative. The authors also present how to automatically generate additively inductive assertions (additive invariants) using Semidefinite Programming (SDP) \cite{Helmberg1999} solvers.

Some efforts focus on automatically generating quantum assertions and adding them to the circuits under test. QuAssert \cite{witharana2023quassertautomaticgenerationquantum}, for instance, addresses the limitations of manual assertion generation, which requires expert knowledge and is inefficient for complex quantum systems. The authors present a method based on static analysis and random sampling to identify key properties of quantum circuits (classical states, superposition, and entanglement). The framework is based on three previously defined types of assertions \cite{HuangStatisticalAssertions, LiuRuntimeAssertions, ZhouProjectionBasedAssertions} to validate different states within the circuit. By automating the generation and placement of these assertions, the framework aims to improve functional coverage while minimizing hardware overhead.

\begin{figure}
	\centering
 \includegraphics[width=\textwidth]{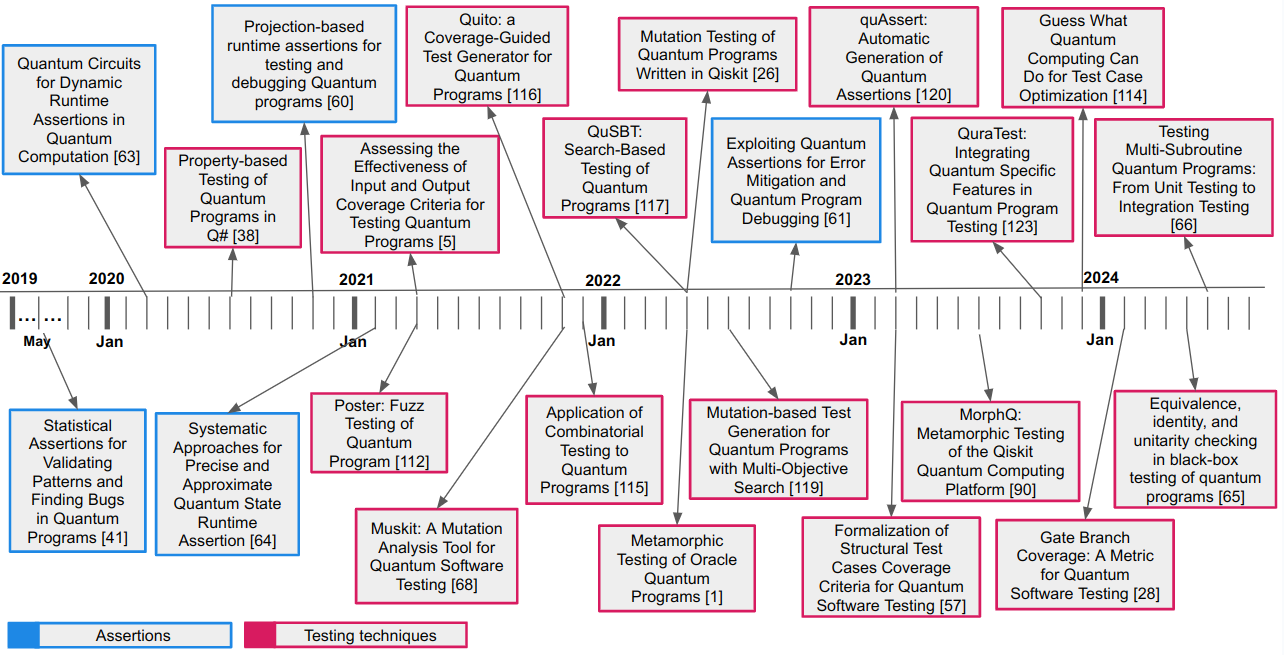}
 \caption{Timeline of studies that address testing techniques and assertions for QPs}
 \label{fig:testing_timeline}
\end{figure}

The timeline of key papers on testing techniques and assertions for QPs is shown in Figure \ref{fig:testing_timeline}. Testing techniques are represented by boxes with red frames and  assertion approaches by box with blue frames.
Based on the timeline, one can note  initial efforts focusing on assertions, in which techniques attempt to solve the challenges from qubit state collapse and the no-cloning theorem. These are followed by works exploring several aspects of software testing. As the field advances, recent research increasingly focuses on leveraging quantum properties, rather than  adapting classical techniques to quantum computing.

\subsection{Bug patterns in quantum programs}
\label{TestingDebuggingQP}

In terms of improving the quality of quantum programs and reducing the likelihood of producing software with defects, there have been efforts focusing mainly on two strategies: (i) studying the most common defect patterns in QPs as well as their characteristics and (ii) developing techniques to automatically identify and fix bugs. Knowing the most recurrent bugs patterns and how they manifest themselves in real QPs can help researchers develop new ways to mitigate these bugs before they occur. This section provides an overview of the bug characterization efforts and the analysis of defect patterns in QPs.

\subsubsection{Bug types and taxonomies}
\leavevmode

Focusing on debugging QPs, \citeauthor{huangMartonosi2018qdb}  \cite{huangMartonosi2018qdb} surveyed a set of quantum computing (QC) algorithms and conducted small-scale experiments. These experiments were based on the implementation and the gradual debugging of each step of these programs. Through this process, the authors identified a set of bug types and proposed methods for mitigating them. The bug types found by the authors are: 
\begin{enumerate}
    \item Incorrect classical input parameters. Errors occur when using classical data as input due to incorrect formats.
    \item Incorrect quantum initial values. Mistakes at initializing qubits to the proper states needed for an algorithm.
    \item Incorrect operations and transformations. Issues arise from using the wrong sequence or type of quantum gates.
    \item Incorrect composition of operations using iteration. Problems with loops or iterations due to limitations in some quantum programming languages.
    \item Incorrect deallocation of qubits. 
    Ancillary qubits remain entangled, affecting the results due to improper deallocation.
    \item Incorrect composition of operations using mirroring. Errors in reversing sequences of operations by applying gates in the wrong order.
    \item Incorrect composition of operations using recursion. Mistakes in implementing controlled operations manually, lead to errors in recursion.
\end{enumerate}

Following a similar structure, but centering their work more on static analysis and less on debugging and testing, \citeauthor{ZhaoIdentifyingBugPatterns} \cite{ZhaoIdentifyingBugPatterns} studied bug patterns in QPs written in Qiskit. The authors identified eight bug types, classified into four areas: initialization, gate operations, measurements, and deallocation. The authors further define bug types for each of the areas listed above. They are:
\begin{enumerate}
        \item {Unequal Classical Bits and Qubits}. This bug type happens when the number of bits defined in the classical registers is smaller than the size of the quantum registers. 
    \item {Custom Gates not Recognized}. In Qiskit, developers can define custom gates by extending the class Gate. According to the authors, this bug type happens mainly when the developer defines a custom gate to act on more than one qubit without considering the basic functioning of composite gates, which consists of combining multiple single-qubit gates and controlled gates to achieve a multiple-qubit gate behavior such as CCX.
    \item {Insufficient Initial Qubits}. This bug type is specific to the use of the TwoLocal class in Qiskit. When defining an instance of such a class, the developer needs to consider the number of qubits needed in the constructor for the parametric circuits or methods involving quantum entanglement.
    \item {Over Repeated Measurement}. This bug type happens when a qubit is measured multiple times, i.e., the result of a quantum register is stored in multiple classical registers. 
    \item {Incorrect Operations after Measurement}. After measuring a qubit, its value collapses to a classical bit. Therefore, it does not hold any qubit properties such as superposition or entanglement. Besides, any qubit entangled with the qubit being measured collapses instantaneously. 
    \item {Unsafely Uncomputation}. This bug type does not affect Qiskit, as qubit deallocation is managed automatically by the framework. However, for certain programming languages, this deallocation process needs to be managed by the developer. When that is the case and qubits are not properly deallocated, an ``unsafely uncomputation'' bug type might happen.
    \item {Inappropriately Modification of Register Size}. This bug type refers to an operation changing the size of a register after it has been defined. This can be a problem for operations involving sets, as the change of the size impacts the hash value defined for that object.
    \item {Method measure\_all}. The measure\_all operation measures automatically all the qubits in a quantum register. This bug type happens when the developer initializes the classical registers and then calls the measure\_all method, making it print the values of the measured qubits and the results used to initialize the classical registers as well. 
\end{enumerate}

In a more general work, \citeauthor{ComprehensiveStudyBugFixes} \cite{ComprehensiveStudyBugFixes} studied 96 real-world bugs and their fixes in four programming languages: Qiskit, Cirq, Q\#, and ProjectQ. The bugs analyzed were collected from public repositories on GitHub and questions posted by programmers in Q\&A websites Stack Overflow and Stack Exchange. The authors found that more than 80\% of the bugs analyzed were related to the quantum-specific parts of the applications. Furthermore, all bugs categorized as having high complexity were found to be associated with quantum computing components in the programs analyzed. 

Following a similar approach as \citeauthor{ComprehensiveStudyBugFixes} \cite{ComprehensiveStudyBugFixes}, \citeauthor{PaltenghiBugsInQP} \cite{PaltenghiBugsInQP} ran an empirical study and analyzed 223 bugs from 18 open-source quantum computing projects, showing that 39.9\% of all analyzed bugs were related to quantum-specific parts of these platforms such as parts that represent, compile, and optimize quantum programming abstractions.
Moreover, the authors show that many of those bugs do not cause the executing program to crash, but return erroneous results, making it more challenging to identify them. The findings indicate that 9.9\% of all bugs analyzed by the authors are related to quantum computing concepts and are placed in components that implement, assemble, and optimize quantum-related routines.

\citeauthor{aoun2022bugcharacteristicsquantumsoftware} \cite{aoun2022bugcharacteristicsquantumsoftware} analyzed bug reports of 125 quantum software projects on GitHub. The authors corroborate previous studies on the complexity of bugs in quantum-related components of the analyzed projects, showing that they are more costly to fix in terms of the amount of code that needs to be changed when compared to classical project bugs. The authors categorized the bugs found in quantum-related components into 13 types:

\begin{enumerate}
    \item {Configuration bugs}. Issues related to configuration files, such as incorrect file paths, directories, or the need to update external libraries.
    \item {Data types and structures bugs}. Bugs involving undefined or mismatched data types and data structures, like data structures with data dimensions or the use of wrong data structures.
    \item {Missing Error handling}. It occurs when exceptions are not properly handled, leading to crashes or wrong behavior.
    \item {Performance bugs}. Problems affecting the stability, speed, or response time of software, including memory issues and infinite loops.
    \item {Permission/deprecation bugs}. Issues related to deprecated calls or APIs, and incorrect API permissions.
    \item {Program anomaly bugs}. Introduced when extending existing code or caused by poor implementations, leading to crashes or incorrect return values.
    \item {Test code-related bugs}. Issues in the test code due to adding or updating test cases, or incorrectly executed tests.
    \item {Database-related bugs}. Bugs concerning the connection between the database and the main application.
    \item {Documentation}. Issues related to typos, outdated documentation, or misleading information in the documentation.
    \item {GUI-related bugs}. Problems with graphical elements like layout issues in text boxes and buttons, which can affect the usability and interpretation of quantum circuits.
    \item {Misuse}. Errors resulting from the incorrect use of functions.
    \item {Network bugs}. Issues related to network connectivity or server problems, including cloud access in quantum components.
    \item {Monitoring}. Bugs due to improper logging, such as incorrect logging levels, excessive logging, or missing log statements.
\end{enumerate}

Extending the work from \citeauthor{PaltenghiBugsInQP} \cite{PaltenghiBugsInQP}, \citeauthor{camara2023fine} \cite{camara2023fine} investigated bugs in 18 open-source quantum computing platforms, splitting bugs between traditional and quantum-specific, analyzing code elements prone to errors as well as the time required for fixes. The study reveals that quantum bugs are typically fixed faster than classical ones and present common code smells in both bug types. The author suggests that, while some classical debugging methods apply to quantum software, debugging techniques that are specific to QC are necessary.

In terms of bugs in QML frameworks, \citeauthor{zhao2023empirical} \cite{zhao2023empirical} investigated 391 real-world bugs collected from 22 open-source repositories of nine popular QML frameworks. They identified that 28\% of the bugs found are on quantum-specific parts of the code, underscoring the importance of developing methods to find and prevent them. The complexity of fixing QML related bugs was a point highlighted in the work by \citeauthor{aoun2022bugcharacteristicsquantumsoftware} \cite{aoun2022bugcharacteristicsquantumsoftware}, as the time to fix them is longer when compared to other bug categories and bugs in classical programs.

\subsubsection{Bug benchmarks}
\label{BugBenchmarks}
\leavevmode

Benchmarks are widely used in classical software testing to simplify repeated experiments and provide repeatable and objective comparisons \cite{Benchmarks}.
Bug benchmarks are datasets with known bugs, usually including the faulty code, a fixed version, and a test to replicate the problem. For Quantum Software, in terms of bug benchmarks, QBugs \cite{QbugsCamposSouto} is a framework that consists of a collection of bugs in quantum software. As QPs  available for testing are scarce, QBugs' authors suggest the creation of an open-source catalog for quantum algorithms, along with a supporting infrastructure that can be used for developers and thus facilitate the execution of controlled empirical experiments.
Following a practical approach, Bugs4Q \cite{Bugs4Q, Bugs4Q2023} is another bug benchmark that consists of 36 real bugs from Qiskit. These bugs were collected, validated, and made available for the community with their respective test cases for reproducing the erroneous behaviors. The idea of the framework is to keep evolving by adding new bugs with new versions of Qiskit, thus building up a reference bug database with their fixes, unit tests to reproduce buggy behavior, and an interface to access and run experiments. 

\begin{figure}
	\centering
		\includegraphics[width=\textwidth]{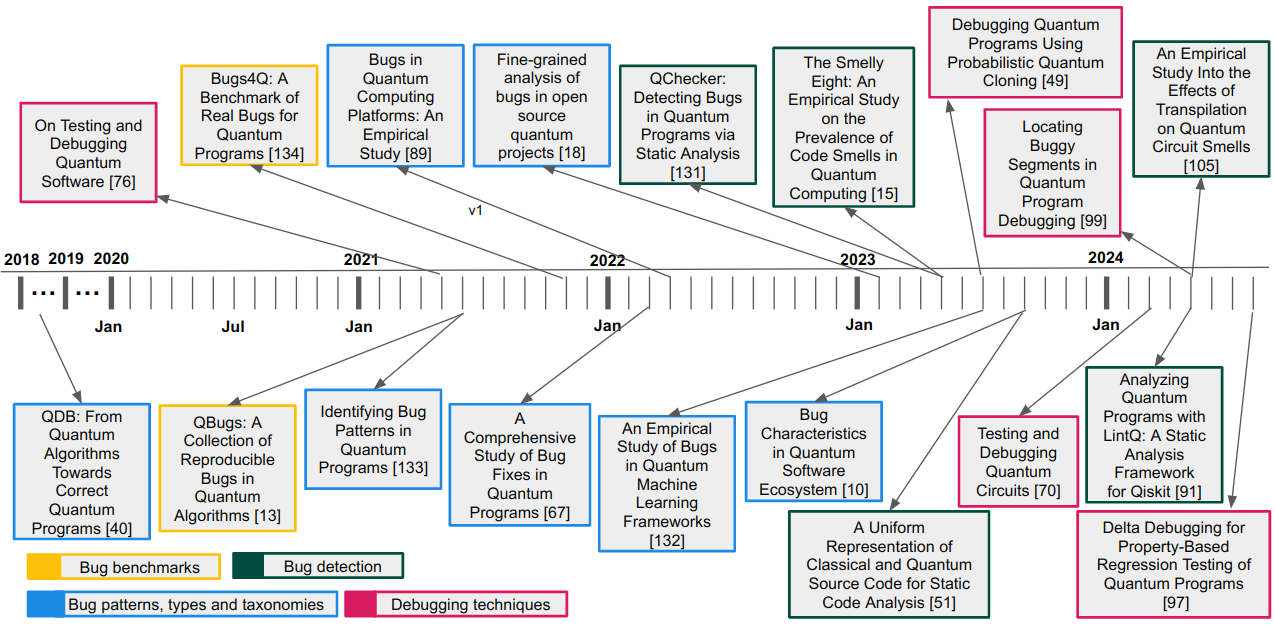}
 \caption{Timeline of studies that address bug taxonomies, types, benchmarks, bug detection and debugging techniques for QPs}
 \label{fig:bug_timeline}
\end{figure}

The timeline of the main papers on bug types and benchmarks for QPs is shown in Figure \ref{fig:bug_timeline} in boxes with blue and yellow frames. It highlights the initial efforts to understand how bugs in QPs differ from their classical counterparts by studying and documenting bug characteristics and patterns. Early works faced the challenge of limited availability of QPs, still an issue in the present day, leading researchers to use open-source frameworks as research subjects to study bugs in QPs. They also implemented small-scale quantum algorithms, such as in the work of \citeauthor{huangMartonosi2018qdb} \cite{huangMartonosi2018qdb}. These studies introduced the first debugging techniques, while parallel efforts, like Bugs Benchmarks \cite{Bugs4Q}, focused on enabling the development and comparison of quantum-specific testing methods.

\subsection{Debugging techniques}

Classical debugging methods such as backtracking, cause elimination, and brute force, have been explored and suggested as possible approaches for debugging quantum programs \cite{OnTestingQPsMiranskyyZhang,MiranskyyQuantumProgramBugFree}. However, debugging quantum programs is currently a challenging problem due to the characteristics of QC such as superposition and the inability to clone quantum states. For instance, a typical debugging approach, which consists of adding print statements in the code to display intermediate values for certain variables, can not be used in QPs due to the collapsing problem. 
Although simulators help to observe quantum states for QPs running on classical computers, they are limited to small programs, as the state explosion for programs with a higher number of qubits makes it unmanageable for classical computers. Furthermore, there are challenges in the interpretation of simulation results, even for small quantum programs \cite{huangMartonosi2018qdb}, which present research opportunities in developing scalable visualizations and improving the interpretability of large-scale graphs that developers can use to inspect and better understand the intermediate states of the QPs being debugged \cite{ThinkingTooClassically}. 

The tooling for developing and debugging quantum algorithms is still limited and has scattered features. Practitioners, for instance, tend to use different tools such as IBM Quantum Composer (a tool offered by IBM that has a circuit composer with a customizable set of tools that allow the developer to build, visualize, and run quantum circuits on quantum hardware \cite{IBMComposer}) and OpenQASM (an imperative programming language designed for near-term quantum computing algorithms and applications \cite{OpenQasm}), alternating between them while debugging quantum algorithms \cite{ThinkingTooClassically}. 
In terms of debugging strategies for QPs, programmers may vary from coarse-grained approaches to fine-grained methods. Coarse-grained techniques are important in scenarios like quantum chemistry simulations in which pair-wise electron interactions lack inherent physical meanings. Fine-grained methods involve inspecting the inner details of intermediate subroutines, allowing comparison of intermediate results with known expected values \cite{huangMartonosi2018qdb}. For instance, \citeauthor{MetwalliToolDebugging} \cite{MetwalliToolDebugging} adapted debugging with slicing (a classical debugging technique) for QPs. The approach consists of dividing quantum circuits into smaller blocks by adding breakpoints in the form of circuit barriers and executing the blocks separately either in a simulator or a quantum computer. The barriers might have the side effect that some qubits are not used in certain slices, thus allowing the user to add a horizontal slice and separate the unused qubits from the analysis. The vertical slicing is similar to the statistical assertions, as the remaining parts of the circuit need to be simulated to allow the inspection of the intermediate states.

The authors created a framework called Cirquo to implement the slicing techniques. In a subsequent paper \cite{MetwalliToolDebugging2}, they define a process to debug quantum circuits and use Cirquo to debug three types of quantum circuits: amplitude–permutation (circuits that permute the amplitudes of quantum states), phase-modulation (circuits which work by changing the phases of quantum states without changing their amplitudes), and amplitude–redistribution circuit blocks (redistribution of amplitudes across several quantum states). Splitting the circuit into smaller blocks for debugging was an approached adopted by \citeauthor{LocatingBuggySatoKatsube} \cite{LocatingBuggySatoKatsube}. The authors developed a method for identifying the location of bugs in QPs by dividing the QP into segments and testing each one individually to locate bugs. A key point in this work is the management of the dependency on the preparation of quantum states from preceding segments with a trade-off between testing accuracy and cost. The proposed method includes a cost-based binary search strategy with early determination for reducing testing cost, and a finalization process for ensuring accuracy.

Another approach focused on the debugging of quantum programs is the work from \citeauthor{Pontolillo2024Delta} \cite{Pontolillo2024Delta}. Their automated debugging technique uses delta debugging, combined with property-based testing, to identify changes in quantum program updates that lead to failures in regression tests. The proposed method was evaluated by injecting both faults and semantic-preserving changes into three quantum algorithms: Quantum Fourier Transform (QFT), Quantum Teleportation (QT), and Quantum Phase Estimation (QPE). The study measures the technique's sensitivity (true positive rate) and specificity (true negative rate), demonstrating high sensitivity and specificity across varying numbers of changes. Sensitivity significantly increased with the number of properties evaluated, while specificity remained consistent regardless of the number of properties and inputs. The results suggest that the method is efficient against both small and large updates to quantum programs and isolates fault-inducing changes, with potential applications in improving the reliability of complex quantum programs. 

There are also debugging strategies that focus on attempting to infer the status of a certain qubit through probabilistic methods. QDebug \cite{Jiang2023}, for instance, allows the developer to set breakpoints in a QP to inspect a certain qubit without collapsing superposition. The approach proposed by the authors consists in using a technique called probabilistic quantum cloning. This technique tries to create a clone of the quantum state being inspected and measuring the cloned state. When the QP reaches the breakpoint, a unitary transformation tries to clone the quantum state being inspected. If it is successfull, the cloned state is measured in order to infer the state of the original qubit. If not, the QP can be rerun until the clone works successfully. This technique is probabilistic and requires the developer to execute the QP multiple times. 

In terms of Automatic Program Repair(APR) for QPs, there are approaches that leverage quantum concepts in their core strategies. \citeauthor{AutomaticRepairQPs} \cite{AutomaticRepairQPs}, for instance, developed UnitAR (UNItary operations Automatic program Repair), a framework that applies unitary operations to repair quantum programs without altering their core qubit evolution. This method uses superposition and entanglement to generate and validate patches automatically.

The efforts discussed in this section show that several advances have been made in testing and debugging of quantum computing. However, there are still gaps to tackle on the road to understanding and proposing effective techniques for the improvement of QPs. The timeline of the main papers about bug detection techniques for QPs is shown in Figure \ref{fig:bug_timeline} in boxes with red frames. In the timeline, one can observe how researchers built on the knowledge about bug characteristics and patterns (Section \ref{TestingDebuggingQP}) to develop  debugging techniques.

\subsection{Program analysis techniques}
\label{subsec:ProgramAnalysys}
The study of bug patterns in QPs serves as the basis for subsequent works that use static and dynamic analysis to identify bugs and code smells in QPs.
In relation to bugs in QPs, QChecker, a framework proposed by \citeauthor{QChecker} \cite{QChecker}, uses static analysis to detect bugs in QPs. The framework is based on eight bug pattern detectors that consist of a refinement of the bugs found in previous works \cite{ZhaoIdentifyingBugPatterns, Bugs4Q}. These bug patterns are: (1) Incorrect uses of quantum gates, (2) Measurement issues, (3) Incorrect initial state (IS), (4) Parameter error (PE), (5) Command misuse (CM), (6) Call error (CE), (7) QASM error (QE), and (8) Discarded orders (DO). The bug pattern detectors are evaluated using the bug patterns defined in Bugs4Q \cite{Bugs4Q}.

Another approach to finding bugs and code smells in QPs using static analysis based on classical Code Property Graphs (CPG) \cite{CPGYamaguchi} is presented in the work by \citeauthor{Kaul_2023} \cite{Kaul_2023}. The authors propose extending the classical CPG framework to include quantum operations, thus allowing the analysis of both classical and quantum parts of a program in a unified model. They propose a tool called QCPG which uses this hybrid approach to detect quantum-related bugs and security vulnerabilities in different quantum programming languages, such as Qiskit and OpenQASM. By extracting the CPG of a program and using a graph database (Neo4J \cite{neo4j}), the tool can detect code smells and bugs such as superfluous operations (quantum gates that do not affect measured qubits), constant classic bits (qubits that are measured but not transformed), constant conditions (if-statements that always evaluate the same), unused result bits (measured but never accessed), and constant result bits (results that never change due to a lack of qubit transformation).

When it comes to code smells, \citeauthor{TheSmellyEight} \cite{TheSmellyEight} derived eight QC-specific code smells from Cirq's best practices \cite{google2024best} and implemented a tool called QSmell that can find these smells in a QP. The smells defined by the authors are (1) Use of Customized Gates, (2) Repeated set of Operations on Circuit, (3) Non-parameterized Circuit, (4) Long Circuit, (5) Intermediate Measurements, (6) Idle Qubits, (7) Initialization of Qubits differently from $\ket{0}$, and (8) No\-alignment between the Logical and the Physical Qubits. The authors validated the tool in a set of Qiskit-based programs and found out that Long Circuit is the  prevalent smell. The smells defined are mostly related to the current hardware limitations and noise of the NISQ era, as they tackle issues with noise propagation, decoherence, or hardware topologies. One can argue that classifying a code practice as a code smell might be subjective. For instance, the smell Intermediate Measurements is seen by IBM not as a bad practice, but as the core of the Dynamic Circuit concept, which might be a way to achieve quantum advantage in the near term \cite{johnson2022dynamic}. The authors use static analysis to identify two of the cited smells, while the remaining six code smells are detected through dynamic analysis, thus requiring the execution of the quantum programs under investigation. 
Before executing the QP in a real QC, the circuit goes through a step called transpilation, which consists of rewriting the input circuit to match the topology of a specific quantum device, and to optimize the circuit for execution on present-day noisy quantum systems \cite{IBMQuantum2024Transpiler}. As such, the transpilation may completely change the original circuit, therefore impacting any metrics related to code smells. 

\citeauthor{Stefano2024} \cite{Stefano2024} investigated the impact of transpilation on quantum-specific code smells and how different target gate sets affect the results. In their study, the authors conducted experiments on 17 open-source quantum programs and 100 synthetic circuits. Their findings showed that transpilation could change the metrics used to detect code smells, even introducing smells into circuits that were previously smell-free. In terms of the changes in the code smell metrics, the authors found that the Long Circuit smell was particularly susceptible to changes during transpilation. Moreover, the choice of gate setting significantly affects the presence and intensity of code smells in transpiled circuits.

Following an approach similar to \citeauthor{QChecker} \cite{QChecker}, \citeauthor{paltenghi2024analyzingLintQ} \cite{paltenghi2024analyzingLintQ} created a framework to detect bugs in QPs that rely purely on static analysis. The tool defines nine analyses that are based on quantum-specific bugs found in previous works \cite{ZhaoIdentifyingBugPatterns}.
The analysis are related to Measurement and Gates (Double Measurement, Operation After Measurement, MeasureAll abuse, Conditional gate without measurement after the associated register, Qubit measured but not transformed), Resource allocation (Insufficient classical bits to measure all qubits and Unused qubits) and Implicit API constraints.

There are also works focused on automatically fixing bugs in QPs such as the work from \citeauthor{AutomaticDiagnosisKher2023} \cite{AutomaticDiagnosisKher2023}. In their paper, the authors propose QDiff, a framework that uses the Python Abstract Syntax Tree (AST) to detect bug-fix patterns in QPs. This framework, further implemented as a proof-of-concept, compares the quantum buggy code and the patches applied to the fixed QP to implement three bug-fix patterns (incorrect initialization of qubits, incorrect gate operations, and incorrect measurements). This work has been extended by \citeauthor{Pranav2023qpac} \cite{Pranav2023qpac}, who developed Q-PAC, a framework for automatic bug-fix pattern detection in QPs. The extended version introduces a research agenda called Q-Repair, aiming to predict fix patterns automatically based on existing bugs. The tool's architecture has been refactored for better extensibility and performance, generating ASTs only once per buggy-fixed code pair and incorporating a coarse filter to reduce the search space. Unlike QDiff, Q-PAC reports all detected patterns for each bug-fix pair and includes detectors for patterns spanning multiple lines.

The timeline of Figure~\ref{fig:bug_timeline} also includes program analysis techniques under the umbrella of bug detection techniques (boxes with green frames). The timeline indicates that these techniques benefited from prior developments on bug patterns, taxonomies, and benchmarks for their own development.
 
\section{A conceptual model for testing and debugging QPs}
\label{sec:conceptual-model}

\begin{figure}
    \centering
    \includegraphics[width=\linewidth]{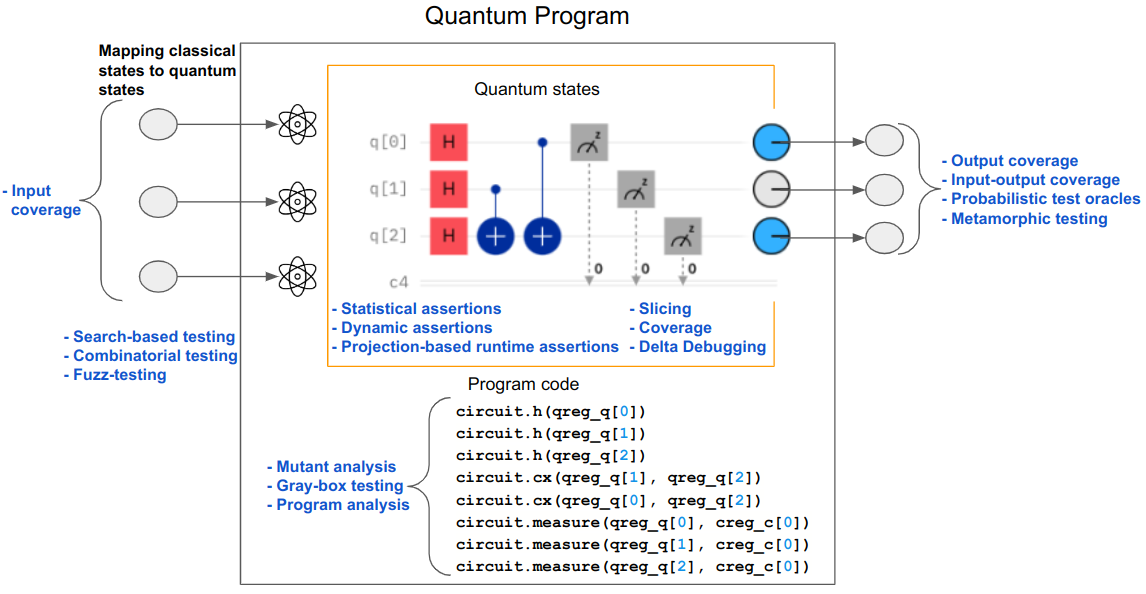}
    \caption{Testing and Debugging techniques in a QP}
    \label{fig:qp_overview}
\end{figure}

In the last section, we made a comprehensive exploration of the state-of-the-art in testing and debugging of QPs. 
Figure \ref{fig:qp_overview} represents a QP and highlights where the reviewed techniques are applied. On the left-hand side, there are the classical data that are mapped into quantum input data handled by QPs. Input coverage, search-based testing data generation, combinatorial testing, and fuzz testing techniques exploit input data to verify the behavior of the QP under test.

The box itself in Figure \ref{fig:qp_overview} represents the QP showing its circuit and code. Metamorphic testing, mutation analysis, gray-box fuzzing, gate and branch coverage utilize the quantum circuit; that is, the QP, to derive testing requirements to exercise it. Likewise, program analysis techniques investigate the QP using static and dynamic strategies to identify bad smells and bug-fix patterns that are associated with or could lead to possible failures. The debugging of QPs requires particular approaches due to QC inherent characteristics (e.g., no-cloning, superposition, entanglement). Because of these characteristics, debugging of QPs remains a complex and limited process. Much effort is being directed to develop ways to assess the QP state without interfering with its behavior by means of assertions. Classical debugging techniques (slicing and delta debugging) have been adapted in the quantum context; though, few tools (IBM Quantum Composer and OpenQASM) are available to practitioners to aid in the debugging task.

The right-hand side of Figure \ref{fig:qp_overview} shows QP's outputs, which can be utilized to derive output and input-output coverage requirements. 
QP's output nature implies that probabilistic oracles are needed. Metamorphic relations can be utilized to check the output and the behavior of the QP. 
Furthermore, noise inherent to current NISQ hardware makes validating the results of QPs a formidable task. The majority of the studies reviewed did not evaluate their strategies and techniques using real quantum computers. As a result, we have few indications that such approaches will be effective when noise and different strategies of transpilation of QPs are taken into account.

\begin{landscape}
\begin{figure}[!htbp]
    \centering
    \includegraphics[width=1.1\linewidth]{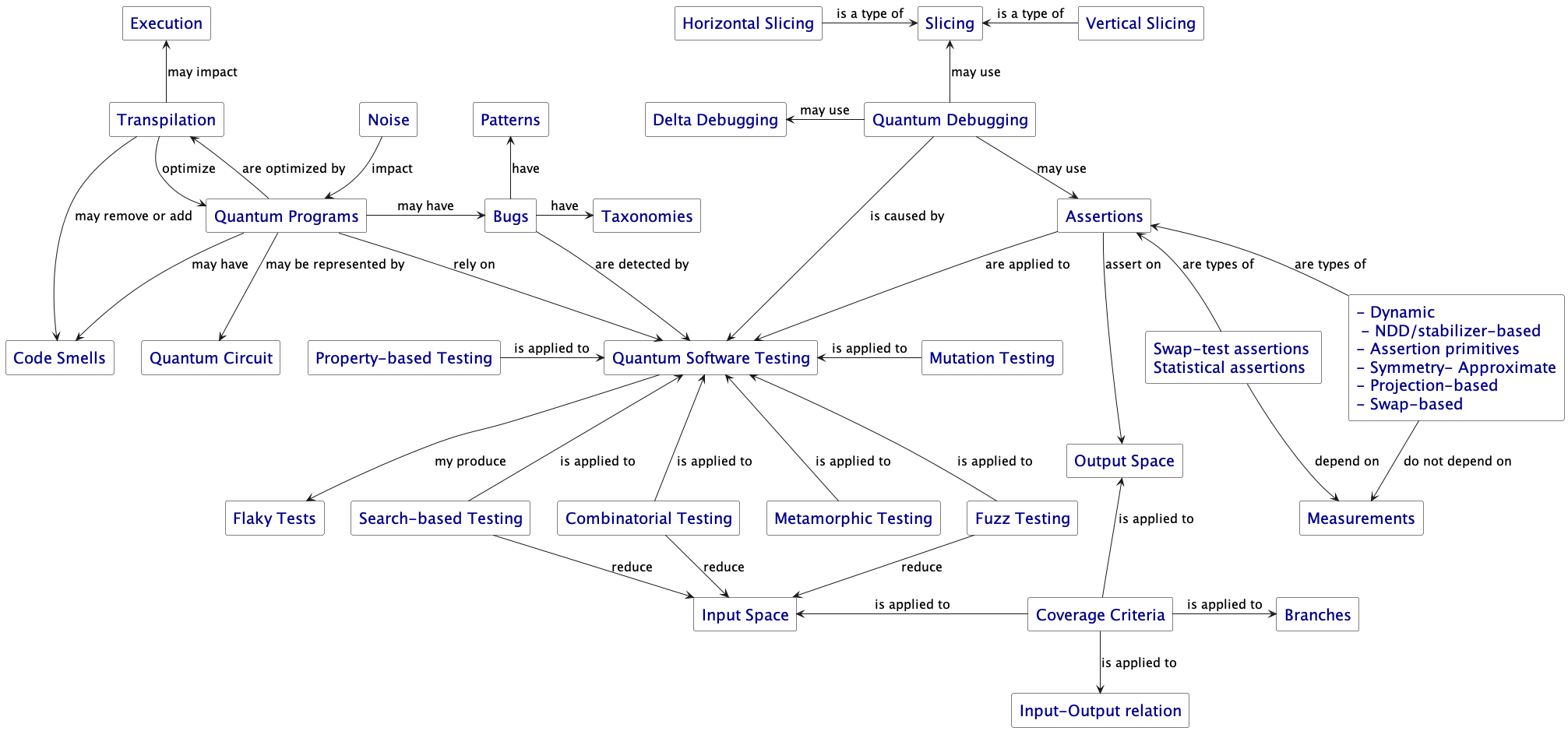}
    \caption{Conceptual model of quantum computing applications}
    \label{fig:conceptual_model}
\end{figure}
\end{landscape}

Figure \ref{fig:conceptual_model} shows a conceptual model
whose goal is to identify the main concepts associated with testing and debugging of QPs and how they are related to each other. QPs may be represented by quantum circuits that execute a sequence of quantum operations on qubits. However, the execution of these programs can be impacted by several factors, including transpilation, which optimizes quantum programs to run more efficiently on specific quantum hardware. Another key challenge is the presence of noise in quantum hardware, which disrupts the quantum state during execution and leads to unreliable results. Noise can degrade the fidelity of quantum circuits, amplifying errors and impacting the overall execution of the program. 

QPs may have bugs, which can lead to incorrect results. To ensure the quality and reliability of these programs, they rely on Quantum Software Testing (QST). Several techniques have been developed for this purpose, including mutation testing, property-based testing, metamorphic testing, fuzz testing, search-based testing, and combinatorial testing. These techniques cover several aspects of a QP illustrated in Figure \ref{fig:qp_overview}, varying from functional to structural approaches. Coverage is also studied in its different forms such as input, output, input-output relation, and branches. 

Assertions are important tools in classical software testing, but they face limitations in the quantum realm as detailed in Section \ref{sec:Assertions}. In QC, assertions can take two main forms: (i) Statistical assertions, which rely on measurements, thus making the quantum state collapse into a classical value; (ii) Dynamic assertions, which allow checks without collapsing the quantum state, helping preserve its quantum properties.

For Quantum Software Debugging, the studies are still in an initial phase. Traditional debugging techniques are difficult to apply in QPs due to the no-cloning theorem and the collapse of quantum states when measured. Slicing (vertical and horizontal) can help in isolating specific parts of the circuit for analysis and debugging. Other approaches tackle the problem with debugging by combining techniques, such as delta debugging and property-based testing. This can help automatically identify defects in QPs by distinguishing between actual defects and semantic-preserving changes. Another important area is the identification of flaky tests, which are tests that produce inconsistent results due to quantum noise or non-determinism in quantum systems.

\clearpage

In the next section, we explore the emerging challenges and opportunities for testing and debugging of QPs.

\section{2030 horizon: Emerging challenges and opportunities}
\label{sec:challenges}

Quantum computing is emerging as an important research area, especially for Software Engineering. The current quantum technologies are evolving, but are still limited in terms of available gates and real quantum computers---the later too expensive for most researchers and practitioners in the current NISQ era. Thus, applications are not exploring the potential benefits of quantum computing. In this scenario, testing and debugging techniques should also evolve to keep the evolution pace concerning the development of high-quality quantum applications for the next years. There are many challenges related to the current state of the field, which is comparable to the early days of classical computing, when programs were written in low-level machine languages \cite{delaBarrera2022}. This lack of higher abstractions to the circuit-based model as well as the absence of quantum-specific testing techniques (unlike the adaptation of their classical counterparts) and tooling can pose extra challenges in the testing and debugging of quantum programs. In this section, we discuss the main challenges and opportunities identified in our research on state-of-the-art testing and debugging techniques for quantum programs, summarized in Table \ref{tab:challenges_summary}.

As previously pointed out by \citeauthor{murillo2024challengesquantumsoftwareengineering} \cite{murillo2024challengesquantumsoftwareengineering}, noise (row 1 in Table \ref{tab:challenges_summary}) is still a problem in the testing of QPs. Most existing testing techniques presented in this paper were developed using error-free simulators, which do not fully capture the realities of quantum noise. However, this assumption does not hold true in the NISQ era when QPs are run in real quantum computers. In terms of error correction, the concept of a ``logical qubit,'' an error-corrected qubit, is a key part of the roadmap for several major industry players, including IBM \cite{IBMQuantumRoadmap}, Google \cite{GoogleQuantum}, and Microsoft \cite{MicrosoftQuantumRoadmap}. Google, in particular, has made recent advancements in error correction \cite{GoogleQuantum}, specifically with schemes such as surface codes, which rely on assemblies of physical qubits to create logical qubits. Though logical qubits are a promising strategy to address current noisy quantum hardware, they imply that more qubits will be needed to tackle already qubit-hungry real problems \cite{gao2022quantum}. As a result, testing techniques capable of checking QPs despite their noisy outputs will be in high demand in the foreseeable future. 

Generally, for QPs, it is difficult to find an oracle and to separate a wrong output from noise in its different forms. For a small subset of QPs that have deterministic output, developing techniques that can separate wrong outputs from errors caused by noise is a promising research direction. Initiatives such as QOIN \cite{MitigatingNoiseMuqeet}---which makes use of machine learning techniques to filter out the noise from outputs---highlight a promising path, though they still face challenges with generalization. Another technique that can help reduce noise, especially for long circuits, is the use of Dynamic Circuits or classical feedforward techniques as implemented in Qiskit \cite{IBMQuantumFeedforward}. These consist of having intermediate measurements in the circuit and applying operations based on the measurement results. Intermediate measurements help reduce the error propagation and accumulation due to decoherence in long circuits. An example of a dynamic circuit with an if statement is illustrated in Figure \ref{fig:dynamic_circuit}.

\begin{figure}[!htbp]
    \begin{center}
        \begin{quantikz}[wire types={q,c}]
            \lstick{$q_0$}     & \gate{H}  &   \meter{} \vcwdouble{1-3}{2-3}{0}  & \gategroup[1,steps=2,style={ultra thick,draw=red!80,rounded
corners,inner xsep=1pt},background,label style={label
position=center,anchor=north,xshift=-0.5cm,yshift=0.1cm}]{if} & \gate[style={fill=blue},label style=white]{X} &  \meter{} \vcwdouble{1-6}{2-6}{0}  &   \\  
            \lstick{$c_0$}    & \qwbundle{1}  &  &  \ctrl{0} \wire[u][1]["c0\_0=0x1"{below,pos=-0.1}]{c} &  &  & 
        \end{quantikz}
    \end{center}
  \caption{A classical feed-forward (dynamic circuit) in Qiskit. \label{fig:dynamic_circuit} Source: \cite{IBMQuantumFeedforward}}
\end{figure}

The circuit begins by applying a Hadamard gate to qubit $q_0$, followed by a measurement. If the measurement result is $\ket{1}$, an $X$ gate is applied to $q_0$, flipping its state back to $\ket{0}$. Then $q_0$ is measured again. As a result, the output of this circuit is always $\ket{0}$. 
This example illustrates a promising technique that can be used in the middle of long circuits to alleviate error propagation, as qubits are measured and the computation can restart with a classical value as the input. Other control flow structures, such as \textit{while} and \textit{for} loops as well as \textit{switch} statements, are also currently supported by Qiskit. 
Another interesting application of dynamic circuits is their use in debugging techniques. Since intermediate measurements can be viewed as a way to modularize the circuit, the programmer can verify certain classical intermediate results for correctness during debugging activities.

When it comes to testing techniques, the issues with handling the combinatorial explosion of the input states in the test of QPs have been well documented \cite{Ali2021Assessing}. This is notorious for pure QPs, i.e., those conceptualized to demonstrate theoretical concepts or to function as small example programs. In practice, a QP will not exist as an isolated entity, but as part of a complete solution with classical components. In this hybrid setup, the first step consists of mapping classical data to quantum data. In these approaches, the inputs of a QP are not only discrete but can also assume continuous values that go through a set of steps to be processed by a QC. Thus, mapping input states with expected outputs and developing testing techniques for the interfaces between the classic and the quantum components becomes even more complex. Following the need to reduce the search space on the input domain, several techniques have been developed such as combinatorial testing, search-based testing, and fuzz testing. These techniques are labor intensive, require significant computational resources and effort, and do not tackle the problem completely, as in most real scenarios the input states are not discrete. Promising research directions (row 2 in Table \ref{tab:challenges_summary}) in this area are: (i) exploring test scalability with realistic data inputs, as noted by \citeauthor{murillo2024challengesquantumsoftwareengineering} \cite{murillo2024challengesquantumsoftwareengineering}, and (ii) the use of quantum input data, such as what has been demonstrated by \citeauthor{PeixunJianjunEquivalence} \cite{PeixunJianjunEquivalence} with Pauli input states (Section \ref{other_testing_approaches}). To build on this, we add the need for testing techniques that use real quantum datasets such as MNISQ \cite{MNISQ} and QDataSet \cite{Perrier2022}. \citeauthor{WhyConsiderQuantumML} \cite{WhyConsiderQuantumML} highlight the current challenges with the lack of truly quantum datasets and the pressing need for standardization of easily preparable quantum states.

On the QP debugging front, one of the major factors impacting future techniques is the inability to clone quantum states (row 3 in Table \ref{tab:challenges_summary}), which completely blocks the inspection of qubit states in the same way that variables are inspected in classical computing. Thus, a simple print statement or the analysis of a qubit in a breakpoint makes the qubit collapse to a classical value, affecting the subsequent execution of the QP. Dynamic assertions and Slicing can be seen as ways to overcome this limitation, albeit with restrictions. For Dynamic Assertions, they can only assert certain characteristics of the quantum state, such as identifying superposition or entanglement. While this approach has practical applications, it shares a limitation with property-based testing:  both focus on verifying properties of the quantum state rather than the quantum state itself. Regarding Slicing, the technique becomes increasingly complex to manage as the size of the circuit grows, making it challenging to handle and debug the individual slices effectively. The use of simulators in the development and debugging of QPs can help to circumvent these issues, but they are limited to small programs due to state explosion. Thus, a variety of classical structural testing techniques (for instance, unit testing with assertions on intermediate results, breakpoints, or print statements for quantum states or memory dumps for quantum state inspection) cannot be directly applied to the testing and debugging of QPs. An interesting approach by NVIDIA \cite{nvidia2021cuquantum} consists in accelerating the simulation of quantum circuits using classical hardware, particularly leveraging Graphics Processing Units (GPUs). They developed cuQuantum, a Software Development Kit (SDK), which provides libraries and tools that allow researchers and developers to simulate quantum algorithms. The idea is to facilitate the exploration of quantum computing techniques before scalable quantum hardware becomes widely available. By using tensor networks and state vector simulation methods, cuQuantum can simulate quantum circuits with tens of qubits. Although GPUs do not fully solve the issues of quantum simulation or the limited access to real quantum computers, they serve as an intermediate step, supporting research and enabling larger simulation tasks until quantum computers become accessible. This technique can be particularly useful for slicing, as certain parts of a QP can be split into smaller blocks that can be simulated by the GPUs.

\begin{figure}[!htbp]
  \centering
  \includegraphics[scale=0.50]{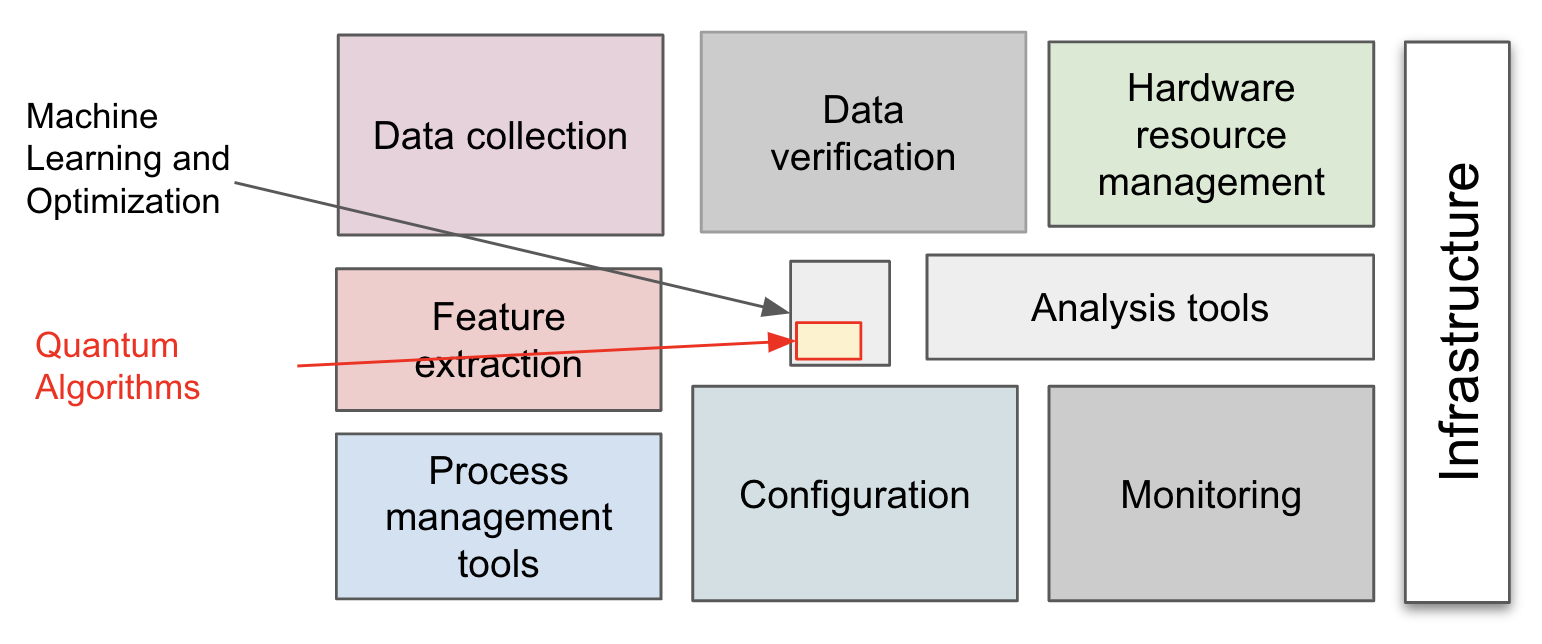}
  \vspace{-2ex}
  \caption{Quantum algorithms in the context of a complete application. Adapted from \cite{GoogleML2022}}
  \label{hybrid_qc}
\end{figure}

Regarding QML applications (row 4 in Table \ref{tab:challenges_summary}), it is observed that, as with other QPs, there are efforts to adapt classical techniques to the quantum space. The similarities between QML algorithms and neural networks make them suitable to share analogous testing approaches. The differences we observe are related to the encoding of the classical values to quantum states as well as the outputs produced by the quantum algorithms to classification labels. Although several techniques cover both topics (such as the encoding techniques discussed in Section \ref{sec:quantum-computing}), the testing strategies are, as far as we know, non-existent. On a broader scope, although some testing strategies exist for pure QC applications, not so much has been developed in terms of the testing of the interfaces between the classical and the quantum world.  As illustrated in Figure \ref{hybrid_qc}, QC applications exist as part of a complete application in which the quantum-specific components will execute a part of the job. Thus, exploring testing alternatives and developing ways to test these interfaces is important. Likewise, understanding bug patterns and their complexity levels in hybrid (classical-quantum) programs---such as the work by \citeauthor{zhao2023empirical} \cite{zhao2023empirical}---, but focusing on real-world applications rather than frameworks, remains an ongoing challenge, especially due to the lack of such applications. As with other quantum programs, hybrid QML applications may present other bug patterns and characteristics that differ from those observed in frameworks. Additionally, adapting classical debugging techniques to quantum programs is a key area that still requires significant development and understanding.

\citeauthor{DiMatteoOnTheNeedEffective} \cite{DiMatteoOnTheNeedEffective} highlights another interesting aspect of the current development practices in QPs, which is the amount of time developers spend integrating different frameworks that follow different styles and conventions (such as order of qubits), types of gates supported, test conventions, and documentation  (row 5 in Table~\ref{tab:challenges_summary}). In such heterogeneous scenarios, converting an algorithm that works correctly in one framework into another might be a challenging task, prone to producing several bugs. Possible challenges may arise from the fact that, as of the writing of this paper, there are no standards defined for a widely accepted quantum computing reference architecture. There is an initiative by the Institute of Electrical and Electronics Engineers (IEEE), with the IEEE P3120 standard \cite{IEEE_P3120_2023}. This is an active project to develop a standard for quantum computing architecture. Approved in 2023 and set to expire in 2025, it aims to define technical architectures for quantum computers, including hardware, low-level software, and programming methods across various qubit modalities (e.g., superconducting, neutral atoms, trapped ions, photonic, NV centers), and mode of operations (e.g., digital and analog). The project is handled by the Quantum Computing Architecture Working Group and is coordinated by the IEEE Computer Society/Microprocessor Standards Committee. It addresses the need for standardization in quantum computing to enhance interoperability and deployment.

Other research directions that have been followed are the study of bug patterns, taxonomies, and techniques for debugging quantum programs   (row 6 in Table~\ref{tab:challenges_summary}). Although bug patterns may address known bugs, helping to reduce their occurrences, there may be other bug types that do not fit into the existing categories. As more studies on bug patterns emerge, they may converge toward a well-defined set of characteristics. This could enable the development of more targeted techniques for handling specific bug categories or debugging QPs with those features. As pointed out by \citeauthor{DiMatteoOnTheNeedEffective} \cite{DiMatteoOnTheNeedEffective}, current studies on bug taxonomies and patterns are based on quantum frameworks and programming languages, which might not be representative of how programmers use them to build QPs. This may be a consequence of the lack of QPs that use these frameworks and programming languages to solve real problems. However, solving real-world problems may require far more qubits than are currently available in today's quantum computers. A report to the US Congress \cite{gao2022quantum} estimates that more than 100,000 qubits are needed to tackle chemical and pharmaceutical simulations. In general, for practical applications, the number of qubits required could reach up to 1 million \cite{europol2024}.

In terms of debugging, although there are scalability issues with the existing approaches, we see a promising path in the developing of techniques that take into consideration strategies for classes of quantum circuits, i.e., circuits that share a certain set of characteristics, similar to what has been proposed for the three circuit types in the work by \citeauthor{MetwalliToolDebugging2} \cite{MetwalliToolDebugging2}. These Quantum Circuit Debugging Patterns (QCDP) may consist of a collection of debugging practices for different types of circuit classes. For instance, Qiskit Algorithms \cite{qiskit2024} define nine categories of algorithms based on the tasks they perform: Amplitude Amplifiers, Amplitude Estimators, Eigensolvers, Gradients, Minimum Eigensolvers, Optimizers, Phase Estimators, State Fidelities, and Time Evolvers  (row 7 in Table~\ref{tab:challenges_summary}).

Another aspect that impacts the application of testing techniques in QPs is transpilation, as it can completely change the original circuit submitted for testing to adapt it to the target architecture. As the circuit being submitted may not be the same executed by the hardware, metrics related to the original circuit may not hold in the final transpiled circuit, as demonstrated by previous works with code smells \cite{TheSmellyEight}. However, depending on the framework being used, the programmer can access the transpiled and optimized version of the circuit being tested. In Qiskit, for example, the transpilation returns a circuit. Therefore, focusing on applying testing techniques to the transpiled circuit may be a promising research direction (row 8 in Table~\ref{tab:challenges_summary}). 

\citeauthor{murillo2024challengesquantumsoftwareengineering} \cite{murillo2024challengesquantumsoftwareengineering} highlight the challenges of creating efficient test oracles for quantum programs. From our point of view, the strategies to reduce the impact of the lack of efficient test oracles will combine a series of techniques such as: (1) Development of testing techniques that can explore properties of the circuits being tested such as metamorphic relations \cite{MetamorphicTestingRuiAbreu, PaltenghiMorphQ} and property-based testing \cite{PropertyBasedHonarvar};  (2) Use of techniques such as Classical Shadowing augmented with machine learning, which do not require as many measures as state tomography, but still infer interesting properties from the quantum state being studied \cite{Huang2020ClassicalShadowing,pennylane2024}; and (3) With the popularization of QML,  switch the interpretation of the test oracle from a "ground truth" with a pass/fail pair to a more generic concept focused on performance metrics as we have in traditional machine learning (e.g. precision, recall, accuracy, and F-score)  (row 9 in Table~\ref{tab:challenges_summary}).

In terms of mutation testing, it is important to investigate how realistic the mutation operators proposed for QPs are; that is, to which extent they can represent real bugs written by a competent programmer (Competent Programmer Hypothesis) \cite{DeMilloCompetentProgrammer}. The studies on bug patterns (Section \ref{TestingDebuggingQP}) provide different lists of bugs identified in quantum-related components of the analyzed programs. However, not all of these bug patterns can be addressed by the mutation operators proposed in existing works on mutation testing. Another significant challenge is the scarcity of real quantum programs (QPs) used in such studies. Most research on mutation testing has been conducted using frameworks like Qiskit or small, textbook examples, rather than on actual QPs. Given the probabilistic nature of both quantum computing (QC) and classical machine learning (ML) and deep learning (DL) systems, several challenges faced by classical ML systems are also relevant to QC systems. For instance, \citeauthor{PanichellaLiemWhatReallyTesting} \cite{PanichellaLiemWhatReallyTesting} emphasize the importance of collaborating with domain experts to identify realistic mutants, faults, and test cases for classical ML and DL programs. This insight is equally applicable to quantum computing (row 10 in Table~\ref{tab:challenges_summary}).

To the best of our knowledge, current works on mutation analysis of QPs focus on first-order mutants; that is, those involving single-point mutations. However, there is a need to study higher-order mutants, which could represent more complex bugs typically found in QPs and help reduce the number of generated equivalent mutants. Similar studies have been conducted for deep learning systems, as shown in \cite{tambon2023mutationtestingdeepreinforcement}. Mutation testing for QPs can also benefit from other testing techniques such as equivalence checking. The challenge of handling equivalent mutants (mutants that behave identically to the original program) is present in mutation testing of QPs as well \cite{ArcainiMuskit, kumar2023development}. Investigating the use of equivalence checking to identify and eliminate these equivalent mutants is a promising research direction, as it could improve the accuracy and efficiency of mutation testing (row 10 in Table~\ref{tab:challenges_summary}).

\begin{figure}
	\centering
		\includegraphics[scale=0.5]{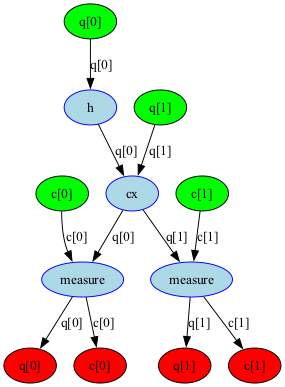}
 \caption{DAG for the Bell states circuit generated by the code in Listing \ref{alg:dag_example}}
 \label{fig:dag_figure}
\end{figure}

Program analysis techniques and QP coverage can be leveraged by exploring the structure of the quantum circuits. A quantum circuit can be represented as a Directed Acyclic Graph (DAG), a structure already implemented in quantum frameworks like Qiskit \cite{qiskit2024}. For example, the method \textit{circuit\_to\_dag} (line 11) from the \textit{qiskit.converters} package allows the DAG for the Bell state circuit in Listing \ref{fig:running_example} to be generated. The corresponding code is shown in Listing \ref{alg:dag_example}, and the resulting DAG is illustrated in Figure \ref{fig:dag_figure}. While DAG representations are primarily used for optimization routines in Qiskit, we envision exploiting them to develop program analysis techniques aiming at identifying bad smells or introducing novel coverage testing approaches. There are similar works in the classical context that use dependence graphs and data-flow analysis \cite{krinkeclonedetection,ChaimBONA23}. In the quantum computing domain, to the best of our knowledge, the only related work is by \citeauthor{Kaul_2023} \cite{Kaul_2023}, which explores a graph-based approach using CPGs for detecting bugs and code smells (row 11 in Table~\ref{tab:challenges_summary}).

\begin{lstlisting}[language=Python, label={alg:dag_example},caption=Code block to produce a DAG from the Bell states circuit]
from qiskit import QuantumCircuit, transpile
from qiskit.converters import circuit_to_dag
import matplotlib.pyplot as plt

qc = QuantumCircuit(2,2)
qc.h(0)
qc.cx(0, 1)
qc.measure(0,0)
qc.measure(1,1)
transpiled_circuit = transpile(qc, optimization_level=0)
dag = circuit_to_dag(transpiled_circuit)
dag.draw(filename='dag_circuit.png')
plt.show()
\end{lstlisting}

The interdisciplinary nature of QC poses an additional challenge to the development of quantum algorithms. On one hand, computer scientists have formal education in software engineering, software testing, and general programming practices, but might lack the necessary understanding of quantum physics \cite{WhenSEMeetsQC} to produce significant contributions to QC. On the other hand, physicists, with a background in quantum mechanics, might lack knowledge of good software development practices \cite{SEChallenges2023}. Moreover, applications for specific domains such as chemistry or economics should count with a diversity of specialties for both the development and evaluation of the resulting outputs. Some authors \cite{QOPAli2023, QAbstractMachines2023} already pointed out the importance of developing quantum-specific paradigms that can abstract away the complexity of working with quantum mechanics concepts. It is also important to incorporate quantum computing concepts into computer science curricula to prepare future computer scientists for this new paradigm. For instance, \citeauthor{Aiello_2021} \cite{Aiello_2021} suggests an interdisciplinary approach to quantum education with the integration of fields like software engineering to prepare a workforce capable of exploring the full potential of quantum computing and developing new techniques across diverse research backgrounds.
This suggestion is confirmed by \citeauthor{Elaoun2021} \cite{Elaoun2021}, who highlight the need for resources to support QP developers. These resources should explain the theory behind specific quantum routines, how to interpret QP outputs, how to fix bugs, and bridge the gap between classical and quantum computing.

A shift in the testing approaches is expected to happen once high-level frameworks and programming languages gain traction. Platforms such as Classiq\footnote{\href{https://www.classiq.io/}{https://www.classiq.io/}} and Silq\footnote{\href{https://silq.ethz.ch/}{https://silq.ethz.ch/}} already explore the idea that QPs can be created with higher abstractions other than the circuit-based model. The argument is that circuit-based quantum programming is the equivalent of creating classical circuits using logical gates such as NAND, OR, NOT, and so on. Although circuits can work for small, simple examples, it does not scale up when it comes to bigger applications. The concept of Quantum Algorithm Design \cite{Minerbi2022} emerges as an attempt to create computer-aided design (CAD) for QPs, in which high-level functional models are created by the user and translated to quantum circuits in the background. These circuits are still subject to transpilation, which introduces its own challenges for testing and verification. For instance, if the generated circuit contains a large number of gates, the results will suffer from noise, as it tends to propagate across the circuit's layers.
As these macro-component approaches mature, there will be challenges testing the interfaces between the usable components proposed by them, as well as with the testing of the individual components themselves. Promising techniques in this area include the use of dynamic circuits in component design, the exploration of functional relationships between inputs and outputs for each component, and the development of techniques for integration testing between components. Similarly to what happened with higher-level programming languages, we expect to see new paradigms being created, as well as design, architectural, and integration patterns for computer systems with both classical and quantum components (row 12 in Table~\ref{tab:challenges_summary}). 

\begin{figure}[!htbp]
  \centering
  \includegraphics[scale=0.60]{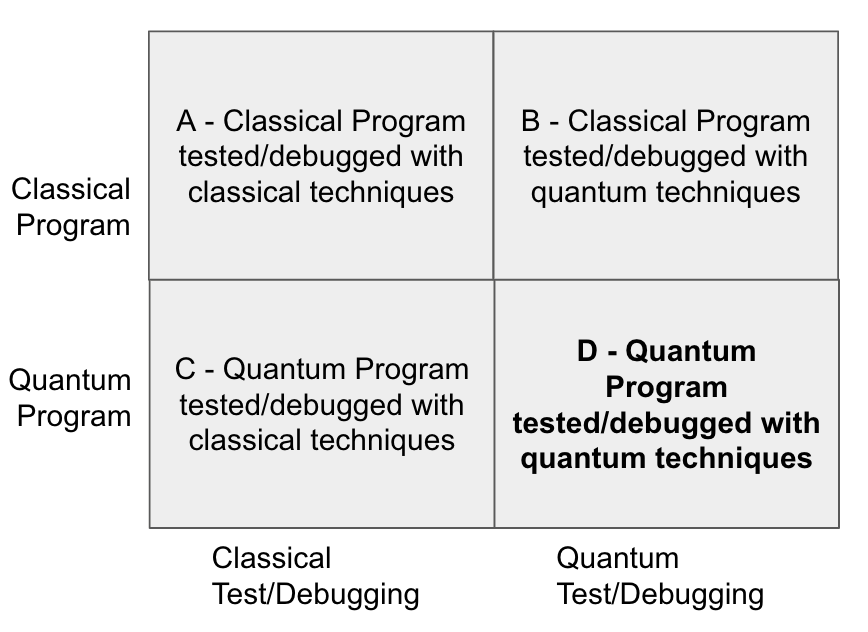}
  \vspace{-2ex}
  \caption{The four dimensions of Quantum Software Testing and Debugging\label{fig:test_matrix}. Based on \cite{Schuld2021Machine} and adapted to the context of Quantum Software Testing and Debugging}
\end{figure}

Figure \ref{fig:test_matrix} summarizes the four dimensions in which testing and debugging of applications fall.
They comprise (1) the type of program being tested or debugged and (2) the type of testing or debugging technique created for a program. In quadrant A, classical programs are tested and debugged using classical techniques. These are the traditional programming, testing, and debugging strategies without any quantum-related elements. In quadrant B, classical programs are tested and debugged with the support of quantum-centered approaches. In this quadrant, the initiatives focus on using quantum-specific phenomena to speed up classical software testing. For instance,  \citeauthor{MetamorphicTestingRuiAbreu}  \cite{MetamorphicTestingRuiAbreu} shown that it is possible to take advantage of quantum parallelism to speed up the testing process of metamorphic rules. Likewise, there have been initiatives \cite{wang2023guess} focused on using QC, more specifically Quantum Approximate Optimization Algorithms (QAOA), in test case optimization problems. Similarly, other works use Quantum Annealers, which are specialized quantum computers for solving combinatorial optimization problems, to tackle the test minimization problem \cite{wang2023test}. 
In quadrant C, a quantum program is tested and debugged with the support of classical testing and debugging strategies adapted to the context of QC. Most of the existing techniques described in this paper focus on this quadrant, as they are about adapting classical software techniques to quantum programs. Although this approach can help us understand the complexities of testing QPs, one should use quantum mechanics characteristics such as parallelism and interference in developing the testing and debugging practices themselves. 
Quadrant D, on the other hand, contains the least explored domain so far, which consists of quantum programs being tested and debugged with quantum-centered approaches. In this case, the tests are developed targeted to the quantum realm and leveraging quantum-specific features such as superposition, entanglement, and interference. Therefore, developing testing and debugging processes that focus on Quadrant D represents a significant challenge and could present valuable research opportunities for the years leading up to 2030.

\begin{table}[htbp]
    \centering
    \renewcommand{\arraystretch}{1.5} 
    \caption{Summary of Challenges and Research Opportunities}
    \label{tab:challenges_summary}
    \begin{tabular}{|p{0.25\textwidth}|p{0.75\textwidth}|}
        \hline
        \textbf{Challenges} & \textbf{Research opportunities} \\
        \hline
        Noise & 
        Development of logical qubits \newline
        Use of ML techniques to identify noise patterns \newline
        Use of dynamic circuits \\
        \hline
        Simplistic input states in quantum testing: focus on basis states $\ket{0}$ and $\ket{1}$ & 
        Create and assess testing techniques using realistic input data \newline
        Use quantum input states instead of basis states \newline
        Use of real quantum datasets \\
        \hline
        Inability to clone a quantum state & 
        Use of Dynamic Assertions to avoid state collapse \newline
        Use of Slicing techniques \newline
        Simulation with support from GPUs \\
        \hline
        Lack of testing and debugging strategies for hybrid QML programs & 
        Develop testing strategies for QML, focusing on testing classical-quantum interfaces (encodings) in hybrid applications \newline
        Investigate and categorize bug patterns in hybrid QML programs, focusing on real-world applications rather than frameworks \\
        \hline
        Lack of a standard for quantum computing architectures &
        Industry to define and adopt a standard \\    
        \hline
        Bug taxonomies and patterns are based on quantum frameworks and programming languages &
        Use as real quantum programs for studies related to bugs \\    
        \hline
        Existing debugging techniques have scalability issues for larger circuits &
        Focus on debugging techniques for classes of quantum circuits in order to create Quantum Debugging Patterns \\
        \hline
        Transpilation may affect the circuit executed by the quantum computer & 
        Develop testing and debugging techniques focused on transpiled circuits \\
        \hline
        Lack of test oracles & 
        Explore circuit and algorithm properties as well as metamorphic testing \newline
        Use of state inspection techniques such as Classical Shadowing \newline
        Adopt performance metrics, as in classical ML systems \\
        \hline
        Mutants for QP are not representative of real bugs & 
        Collaborate with domain experts to design mutants closer to real bugs \newline
        Develop high-order mutants to capture the complexity of real bugs \newline
        Use equivalence checking to reduce the number of equivalent mutants \\
        \hline
        Program analysis and coverage & 
        Explore circuit representations based on graph-based structures \\
        \hline
        Need of high-level abstractions for quantum programming, testing, and debugging & 
        Develop high-level abstractions for quantum circuits to avoid low-level complexity \newline
        Explore input-output relations and dynamic circuits in the test of components and their interfaces
        \\
        \hline
    \end{tabular}
\end{table}

\section{Related Work}
\label{sec:related-work}

In terms of testing QPs, there are systematic studies \cite{GarciaQSTStateArt,QSE_Landscapes_Horizons, delaBarrera2022, fortunato2024verification} that present a broad overview of different techniques, varying from adapting classical approaches to QPs with multiple measurements and statistical analyses to using Hoare logic to determine whether a quantum program is correct. These works also cover topics such as Bug Benchmark frameworks for QPs \cite{QbugsCamposSouto, ZhaoIdentifyingBugPatterns}, reusability of quantum algorithms, data structures, and libraries \cite{OffTheShelfComponentsforQP}, quantum algorithm development \cite{Steiger_2018}, formal verification of quantum protocols \cite{Anticoli2018}, simulation of QPs \cite{Smelyanskiy2016}, and other traditional classical methods such as Mutation Testing, Property-based testing, and Fuzz testing. Other authors \cite{QSE_Landscapes_Horizons} explore the issues with testing and debugging QPs from a different perspective: they study bug types, taxonomies, bug repositories \cite{QbugsCamposSouto, Bugs4Q}, and benchmarks \cite{huangMartonosi2018qdb,  ZhaoIdentifyingBugPatterns, ComprehensiveStudyBugFixes}. 
These systematic reviews cover other topics such as Assertions types \cite{YingInvariants, HuangStatisticalAssertions, ZhouProjectionBasedAssertions, LiuRuntimeAssertions}, and the overall challenges associated with the testing of QPs \cite{OnTestingQPsMiranskyyZhang, miranskyy2021testingdebuggingquantumsoftware}.

Similarly, \citeauthor{murillo2024challengesquantumsoftwareengineering} \cite{murillo2024challengesquantumsoftwareengineering} propose a roadmap for 2030 with research topics in QC in the following areas: (1) Service-Oriented Computing, (2) Model-Driven Engineering, (3) Testing and Debugging, (4) Programming Paradigms, (5) Software Architecture, and (6) Software Development Processes. Concerning testing and debugging QPs, the authors highlight several challenges. First, they discuss the need for efficient test oracles, as current approaches such as statistical assertions, multiple executions, and projection-based runtime assertions are computationally expensive and require program specifications. Second, the authors point out that test scalability is limited, as existing strategies for generating test data are simplistic and focus on base states like \( \ket{0} \) and \( \ket{1} \). Another issue is the gap between simulators and real quantum computers, as current test methods rely heavily on simulators and do not account for the noise present in actual quantum devices. Lastly, the authors address the further development of testing methods, emphasizing that mutations used in mutation testing might not accurately represent real faults, and a large number of mutants are often generated. 

As the testing and debugging approaches evolve and progress is made on different fronts, the ideas and tools implemented by them should be usable and easily reproducible by researchers and practitioners. \citeauthor{fortunato2024verification}~\cite{fortunato2024verification} presents an overview of current testing approaches and analyzed 16 recent quantum software testing techniques focused on quantum circuits. They applied the techniques with available tools to the Bell State QP written in Qiskit, highlighting challenges in the setup of each tool and noting that some proposed ideas lacked implementations. The authors pointed out the lack of quantum fault benchmarks, which could facilitate the evaluation, comparison, and reproducibility of different techniques proposed.

When compared to the papers previously cited, our work goes deeper into the software testing and debugging aspects in QC by making a thorough characterization of the state-of-the-art with the  discussion of  fundamental and recent works   and by identifying 12 challenges and their respective research opportunities. 
Furthermore, we developed a conceptual model that illustrates the relationships between key concepts in quantum testing and debugging and helps locate the current challenges and opportunities in the field. Topics such as noise, characteristics of the NISQ era, transpilation, and the lack of a standardized quantum computing architecture are explored with a focus on their impact on the application of the testing and debugging techniques being developed. We also discuss the interfaces between classical and quantum systems, especially in QML applications, and the existing approaches for converting classical data into quantum states, highlighting the need for testing techniques focused on these interactions. Moreover, we emphasize the importance of interdisciplinary collaboration and investment in education programs on quantum software engineering, in general, and testing and debugging, in special, given the interdisciplinary nature of quantum computing.

\section{Conclusions}
\label{sec:conclusions}

In recent years, Quantum Computing has emerged as a promising field due to its capabilities of solving complex problems and the developments in quantum hardware. As we approach 2030, the rising interest in quantum programming languages and frameworks underscores the importance of studying and developing specialized techniques for testing and debugging quantum programs. 
In this paper, we presented an overview of the main concepts and techniques for testing and debugging quantum computing applications, illustrated their relations in a conceptual model, and identified 12 challenges along with research directions to address them. 

For the testing of QPs, we highlighted the characteristics and limitations of quantum computing and their impact on the development of testing strategies. These include noise, the no-cloning theorem, and the lack of standardization in terms of a reference architecture for quantum computers. We explored the challenges and opportunities in areas such as transpilation, mutation analysis, testing interfaces between classical and quantum domains in hybrid applications (commonly implemented as QML solutions), program analysis, and defining coverage criteria for different parts of a QP. In terms of debugging, we presented strategies ranging from classical techniques adapted for QPs to more advanced ones such as slicing and delta debugging with property-based testing, which leverage aspects like entanglement and superposition. We also presented works on the classification of bug patterns in quantum programs, particularly in QML applications. These initiatives are summarized on two timelines: one illustrating the evolution of testing techniques for QPs over the past six years, and another detailing advancements in bug taxonomies, types, benchmarks, bug detection, and debugging techniques.

We proposed research opportunities for each of the 12 challenges we identified across the different studies. These opportunities address some of the fundamental aspects of quantum computing, such as developing strategies to handle noise (e.g., using machine learning to identify noise patterns or dynamic circuits to mitigate noise propagation), as well as specific characteristics of certain testing techniques, such as using mutants that reflect realistic bugs in mutation analysis.

For debugging, we discussed the scalability issues in existing strategies and proposed potential research directions, including the use of slicing techniques supported by GPUs and the development of debugging methods for specific classes of quantum circuits. These challenges and opportunities were summarized in a table for clarity and to organize the discussion.

The research directions presented here intend to guide software engineering researchers interested in quantum computing, helping them develop new approaches to address the challenges of testing and debugging quantum programs.

\bibliographystyle{ACM-Reference-Format}
\bibliography{main}

\end{document}